\documentclass[final,3p,times, 12pt]{elsarticle}

\usepackage{amssymb}
\usepackage{amsmath}
\usepackage{import}
\usepackage{graphics}
\usepackage{graphicx}
\usepackage{svg}
\usepackage{caption}
\usepackage{subcaption}
\usepackage{algorithm}
\usepackage{empheq}
\usepackage{xcolor}
\usepackage{eqparbox}
\usepackage{booktabs}
\usepackage{showlabels}
\journal{Journal of Fluids and Structures}

\begin{document}

\begin{frontmatter}



\title{Numerical investigation of fluid-structure interaction in a pilot-operated microfluidic valve}



            
\author[inst1,inst2]{Ahmed Aissa-Berraies}

\affiliation[inst1]{organization={Computational Mechanics \& Advanced Materials Group, Department of Civil Engineering and Architecture, Università degli Studi di Pavia},
            addressline={Via Adolfo Ferrata}, 
            city={Pavia},
            postcode={27100}, 
            country={Italy}}
\affiliation[inst2]{organization={Faculty of Mechanical Engineering, Multiscale Engineering Fluid Dynamics Section},
            addressline={P.O. Box 513}, 
            city={Eindhoven},
            postcode={5600 MB}, 
            country={The Netherlands}}  
\author[inst2]{E. Harald van Brummelen}
\author[inst1]{Ferdinando Auricchio}

\begin{abstract}
The present paper is concerned with numerical investigation of the performance of a pilot-operated control valve based on shape memory alloy actuation control. The valve under investigation can be integrated into miniaturized hydraulic systems and is developed to perform precise dispensing, mixing, or dosing tasks while being able to withstand relatively high pressure differences. The study evaluates the valve's response under the current ON/OFF and the desired proportional control regimes using numerical methods for fluid-structure interaction. The computational model replicates the operation of the valve, which requires an understanding of the complex interactions between the fluid flow with the pressurized valve and the contact with the valve seat during the opening and closing processes. In addition, the model leverages advanced numerical techniques to overcome several complexities arising mainly from the geometrical, material, and contact nonlinearities, and to mitigate the shortcomings of the partitioned fluid-structure interaction approach. Several simulations are conducted to examine the valve's structural and flow behavior under varying pressure conditions. Results indicate that the valve is adequate for ON/OFF actuation control but is susceptible to flow-induced vibrations during the proportional control regime that occurs due to the sharp pressure drop in the valve-seat gap and the ensuing Venturi effect, which counteract the opening of the main valve. The fluid-structure-interaction simulations provide insight into the mechanism underlying the flow-induced vibrations, which can serve to improve the design and enhance the performance of the valve in microfluidic applications. 
\end{abstract}

\begin{graphicalabstract}
\includegraphics[width=\textwidth]{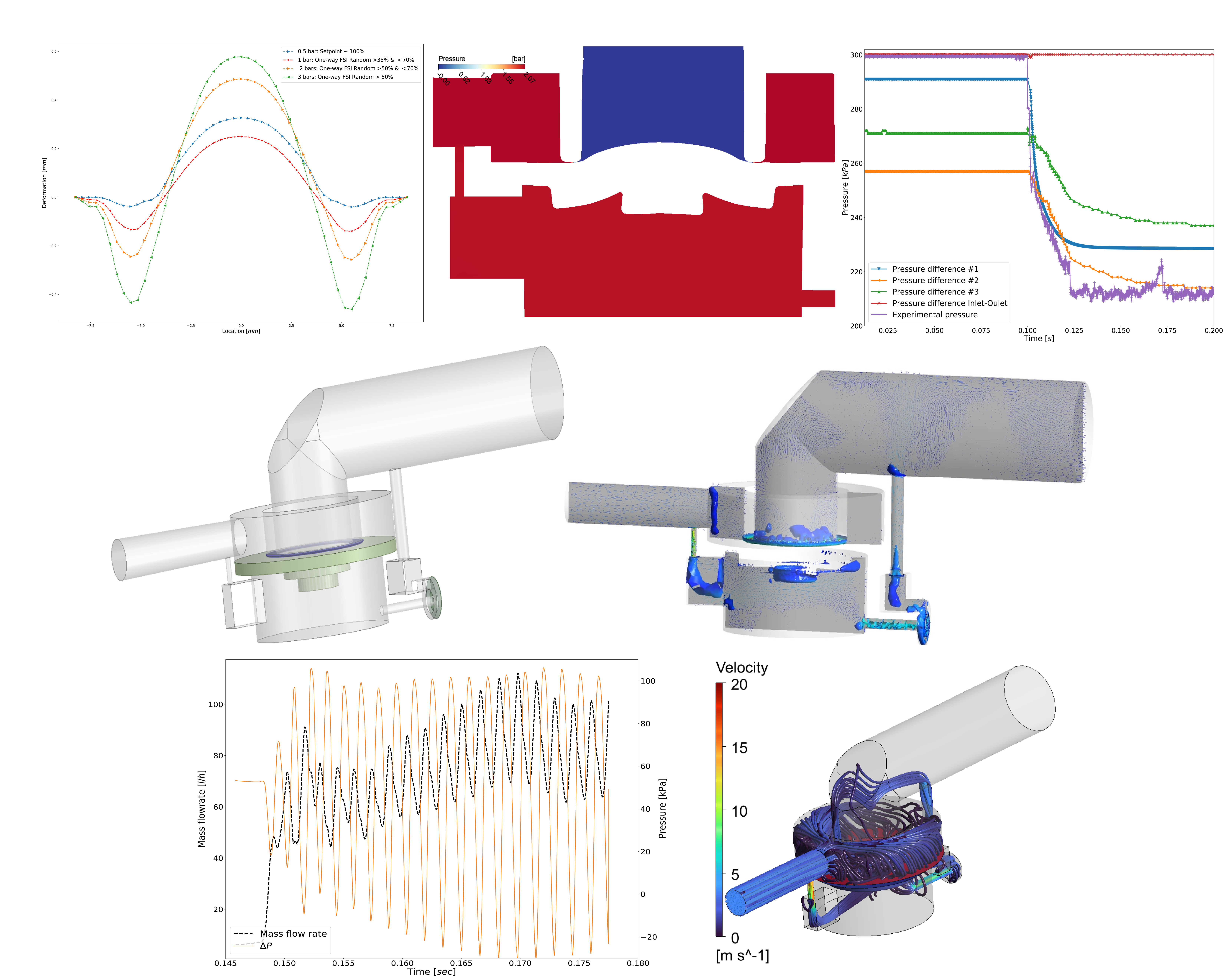}
\end{graphicalabstract}

\begin{highlights}
\item Numerical modeling of a microfluidic pilot-operated valve using partitioned fluid-structure interaction methods
\item Solution strategy for the multiphysics problem, taking into account high added-mass effects and contact events
\item Valve performance evaluation under ON/OFF and proportional operation regimes
\item Assessment of flow-induced vibration during partial opening
\end{highlights}


\end{frontmatter}


\newcommand{\EQ}[1]{(\ref{eq:#1})}
\section{Introduction}
\label{intro}
During the last four decades, the downscaling of piping systems has led to the miniaturization of mechanical devices like valves and pumps. Extensive research projects have been conducted to reduce the size of hydraulic devices to the mesoscale and microscale levels. This trend has enabled the commercialization of smaller products, such as coffee machines, dosing systems, and dispensing machines, endowed with efficient and sophisticated hydraulic circuits. Likewise, the leverage of microfluidic devices allowed for advanced scientific development in medical research. As an example, it supported the study of RNA sequencing \cite{Sarma2019} and intercellular interaction \cite{Pang2021}, which has eventually accelerated the development process of various medicines and drugs, such as the COVID-19 vaccines. 

Valves are designed to be integrated into a circuit and control the fluid flow, the pressure, or the flow direction. Several valve types are used in various industrial applications since they can execute one or multiple services, such as starting and stopping the fluid flow, flow throttling, pressure regulation, relieving the circuit from overpressure, and redirecting the flow. However, it may become challenging to accommodate multiple valves with different features as required to deliver a specific flow service, especially in miniaturized hydraulic systems. Such a limitation has encouraged the realization of multi-tasking microfluidic valves that can accomplish two or more functions at the same time. In this context, some devices can start, stop, and change the flow direction, while others can control the flow and flow direction and simultaneously regulate the fluid flow rate. Various miniaturized valves have been successfully developed and commercialized \cite{Oh2006}. In particular, the pilot-operated dual valve is one of the prominent types nowadays. Such a valve can be perceived as modular, i.e., with two stages mounted together to form a single unit. The pilot stage controls and regulates the flow, while the main stage controls the direction of the flow. The device is designed to improve the response time, reduce the power consumption compared to single-stage valves, and ultimately relax the hydraulic circuit density~\cite{czechowicz2015shape}. 

This paper presents a computational investigation of a smart pilot-operated control valve (POCV), i.e., a two-way medium-separated valve comprising two `normally closed' diaphragm-type stage valves \cite{Natarajan2017}. The POCV is designed to operate in two modes: The first mode is where the valve operates as an ON/OFF control valve. That is, the actuator shifts into an open or closed position as a function of a specific digital electronic input. The second mode regards the POCV operating in a proportional control regime, where the actuator moves to any intermediate position corresponding to a varying electronic (analog) signal. The POCV is smart as it operates using a shape memory alloy-based actuator controlled by an electronic system that can be connected to cloud and mobile platforms for remote data exchange, management, and diagnostics. The shape memory alloy (SMA) wire is a soft, lightweight, and highly flexible material that can recover from different temporary shapes to its initial shape under temperature control variations \cite{Giulia_scalet}. Ultimately, the microfluidic device aims to provide a precise throughput and handle high-pressure differences compared to other microfluidic valves of similar size. 

Though the POCV has proven effective in controlling the flow for the ON/OFF mode, limitations during the proportional operating mode were noticed. Experiments during the proportional opening process revealed the emergence of severe vibrations as soon as the actuation reaches a specific position. These vibrations are associated with noise and turbulent outflow and are undesirable for the performance of the valve. Several trial-and-error procedures were conducted to mitigate the problem, but without success. The inability to address these issues is mainly attributed to the miniature size limitations, and the lack of advanced equipment to precisely locate these vibrations.

Recent advances in numerical methods have shown promising prospects for industrial decision-making on complex production technologies supported by simulation-based engineering \cite{protechtion}. Specifically, computational simulations have been employed to investigate valve dynamics, providing a deeper understanding of mechanical behavior and giving insights into performance within industrial processes. For instance, a $3$D check valve model is developed and tested using numerical methods for fluid-structure interaction (FSI) in \cite{Alberto2019} to analyze the sealing characteristics of the diaphragm volumetric pumps. A numerical tool using computational fluid dynamics (CFD) is developed in \cite{Beune2009} to compute the mass flow capacities and opening characteristics of spring-loaded safety valves at operating pressures up to $3600$ bar. The study is later extended in \cite{Beune2012} to cover FSI numerical methods to analyze the opening characteristic and eventual instability issues to be avoided. On the other hand, several studies focus on analyzing the valve flow response as a function of membrane deflection at mesoscale and microscale levels. An FSI simulation is performed in \cite{Lin2020} to evaluate the impact of the membrane material properties on the flow behavior in a microvalve. A microfluidic device with a single micro-channel is fabricated and tested in \cite{Chimakurthi2018} using FSI computational methods to investigate the fluid interaction with the deformable wall. The work of \cite{natarajan2017analysis} exemplifies the significance of the structural material properties and the topology of the inlet region in the microvalve performance. Furthermore, a few studies focus on the pilot-operated valve performance: A dynamic model is developed in~\cite{Wei2018} to analyze the bifurcation within a hydraulic system composed of a pilot-operated chamber valve and a piston. The analysis shows the conditions leading to instability and provides a guideline to initiate an optimal design. A simulation tool is developed by \cite{Allison2016} and consists of a $1$D fluid model coupled with a structural piston to predict the instability of a pilot-operated pressure relief valve. The results were validated after a confrontation with experimental data. Nevertheless, it is interesting to mention that only scant attention has been given to pilot-operated valves at the microfluidic scale. Although a pilot-operated miniature digital hydraulic valve with a high flow capacity and a fast response was introduced in \cite{Lantela2014}, authors are unaware of additional research studies covering computational simulations of microfluidic pilot-operated control valves, let alone instability issues at smaller scales. This can probably be attributed to the computational challenges of modeling such problems using multiphysics simulation tools, the limited knowledge about these devices' working operations, or exclusive innovation copyrights in designing and developing these products to operate within microfluidic circuits.

The novelty in the present contribution pertains to leveraging advanced simulation capabilities to overcome experimental limitations and understand the performance and dynamics of the microfluidic pilot-operated valve under investigation. In particular, numerical FSI methods are employed in this work to reproduce the interaction of the fluid and the structural membranes during the valve ON--OFF and proportional regimes. The latter includes the onset of flow-induced vibrations. 

Numerical methods for FSI have gained significant interest over the past two decades as an analysis instrument in various engineering and scientific fields, e.g. aerospace and civil engineering, transportation, medical applications, and the beverage industry. Several applications are concerned with multiphysics problems that can be solved using numerical FSI methods, such as the motion of an aircraft's wings due to the actions of the air at high-speed~\cite{Saeedi}, the fluid's effect on valves during operational conditions~\cite{Ogawa1995,Lin2020,Liu2019}, airbag deployment for passenger safety in vehicles~\cite{timo_thesis}, and the investigation of cardiovascular disorders~\cite{Radtke2017,cai2019some}. Although modeling the transient behavior of a moving structure interacting with a fluid remains a challenging task, considerable progress has been achieved following the advancements in computational fluid dynamics (CFD), computational solid mechanics (CSM), and high-performance computing (HPC).

Numerical methods for FSI can generally be categorized into monolithic and partitioned approaches. Monolithic methods have been propounded as the more robust~\cite{Michler2004}. However, these methods are in fact abstruse in view of the need to develop dedicated codes. In addition, the matrix system of equations is generally dense and ill-conditioned~\cite{Richter:2015kq} and, hence, monolithic methods can be memory 
expensive~\cite{chimakurthi2018ANSYS}. Partitioned approaches, on the other hand, provide more versatility in compromising between robustness and efficiency~\cite{van2011partitioned}. Moreover, and in practice most relevant, partitioned methods preserve the modularity of the fluid and solid subsystems, thus enabling the reuse of advanced CFM and CFD software, e.g. commercially available software such as ANSYS (ANSYS Inc., Canonsburg, PA, USA)~\cite{ANSYS2021} or open-source software such as OpenFOAM. The advantages of modularity do not just pertain to the software, but also to the computational models that have been developed, often over periods of many years, for computational structural and flow analyses~\cite{chimakurthi2018ANSYS}.

To assess the suitability of partitioned FSI approaches based on commercially available software for complex FSI problems such as the considered valve system, in the present contribution we regard the use of the ANSYS simulation software. The valve system under consideration carries two essential complications for partitioned methods. First, the fluid is incompressible and has a relatively large density relative to the membrane material. As a result, the fluid carries a relatively large added mass which, owing to the incompressibility, cannot be controlled by means of the time-step size~\cite{causin2005added,van2009added}. This means that the FSI problem is intrinsically strongly coupled. Accordingly, a strongly coupled partitioned approach must be applied, in which the fluid and solid subsystems are solved several times within each time step. This is to be contrasted to weakly coupled FSI problems, which can be solved utilizing a so-called staggered method, in which the fluid and solid are solved only once per time step~\cite{Farhat2010}. Strongly coupled partitioned methods are generally based on a simple fixed-point iteration approach referred to as {\em subiteration}~\cite{van2011partitioned}. Subiteration has proven successful for, for instance, 
butterfly-valve-actuation simulations~\cite{Ezkurra2018}. Within each time step, the FSI solution obtained by subiteration in principle coincides with that of a monolithic approach. 
However, the robustness and efficiency of subiteration deteriorate as the inertial effects of the fluid on the structure become more dominant, i.e., as the FSI coupling strength as expressed by the fluid-solid mass ratio increases. In cases where the fluid-solid mass ratio is too large, subiteration is unstable or converges prohibitively slowly. In such cases, auxiliary techniques are required to enhance the subiteration convergence behavior, e.g. artificial compressibility (AC)~\cite{degrooteAC}, Aitken's $\Delta^2$ method~\cite{Mok}, or the interface quasi-Newton inverse least squares (IQN-ILS) methods~\cite{DEGROOTE2009793,HAELTERMAN20169}; see~\cite{van2011partitioned} for an overview.

The second fundamental complication of the considered valve FSI problem in the context of partitioned solvers is that 
in the closed configuration of the valve, the fluid volume upstream of the flexible membrane is fully closed. The standard partitioned approach for FSI relies on a so-called Dirichlet--Neumann partition of the FSI interface conditions, in which the fluid tractions are imposed on the solid as a Neumann (natural) boundary condition, and the solid displacement is imposed on the fluid as a Dirichlet (essential) boundary condition. If the valve is closed, the upstream fluid volume is subjected to Dirichlet boundary conditions on its entire perimeter, which, in combination with the incompressibility of the fluid, leads to a compatibility condition on the solid deformation. This problem is in the FSI literature referred to as the {\em incompressibility dilemma}~\cite{Kuttler_incompress}. It is to be noted that in the valve problem, also nearly closed configurations occur, which carry similar problems. The compatibility condition underlying the incompressibility dilemma can be avoided by reverting to a Robin--Neumann (or Robin--Robin) partitioning, in which the fluid is subjected to a mixed-type boundary condition at the fluid-solid interface~\cite{Badia2008,Badia2009,Fernandez:2013qf}. However, this partitioning strategy is typically unavailable in black-box FSI solution procedures. Another approach to circumvent the compatibility condition is to introduce artificial compressibility of the fluid in the iterative procedure~\cite{bogaers2016evaluation}, or simply assume compressibility of the fluid in the FSI model~\cite{laspina}.

The numerical investigations in this paper focus on the performance of the considered POCV valve for digital and analog signal inputs, i.e., in standard ON/OFF operation and in a proportional-control setting. The computational model takes into account the SMA-based actuator displacement acting on the pilot valve and the main valve sealing properties over time. To account for the sealing of the valve, in addition to the interaction of the fluid and solid subsystems, the considered valve model also incorporates the contact between the membrane and the valve seat and, hence, in fact corresponds to a fluid-solid-contact interaction problem~\cite{Ager:2021rf}. The fluid and structural models (incl. contact) are simulated with the ANSYS Fluent and ANSYS Transient Mechanical modules, respectively. The data mapping across each solver is handled via ANSYS SystemCoupling. An advantage of the ANSYS software, highlighting the benefits of reusing software modules that are enabled by the partitioned FSI approach, is that it facilitates advanced geometrical modeling of the valve system, provides functionality to handle contact, gives access to nonlinear material models and means of calibrating such models, and provides access to turbulence models. In addition, as the numerical model of the valve system generally involves many degrees of freedom, one can exploit the various parallel computing capabilities offered by ANSYS.

The remainder of the paper is organized as follows: Section~\ref{sec:POCV} introduces the microfluidic pilot-operated control valve and provides its characteristics and specifications. The mathematical-physical model of the valve system and its numerical approximation are elaborated in Section~\ref{sec:ModMet}. This section also considers the utilized coupling strategy. Section~\ref{sec:NumRes} presents the numerical experiments and discusses the results. Finally, Section~\ref{sec:Concl} presents concluding remarks.

\section{Pilot-operated control valve: Characteristics and operating conditions}
\label{sec:POCV}
The valve under investigation is a smart pilot-operated control valve (POCV) developed by a high-tech start-up that designs and develops smart and precision microfluidic systems. This smart valve is designed to provide precise flow regulation and positioning accuracy during operation, with ON/OFF and proportional control working regimes that can be effectively integrated into hydraulic circuits for medical, food, and beverage applications. In addition, the device is equipped with flow, pressure, or temperature sensors, allowing closed-loop control of the customer's desired function, such as autonomous dosing, dispensing, profiling, or mixing, based on the signal from the connected sensor. In the current state of the art, the control is performed by the movement of a pilot membrane, activated by a shape memory alloy-based actuator that changes the displacement of the moving part, which enables an ON/OFF actuation with two positions (open or closed), or a proportional actuation that can assume intermediate positions.

The valve is modular and consists of two medium-separated chambers, \textit{viz.} the main chamber and the pilot chamber.  The pilot chamber part is connected to an electronic circuit through a miniaturized shape memory alloy (SMA) material wire. This smart material can change its shape under temperature variations~\cite{czechowicz2015shape,Giulia_scalet}. The activated SMA material shape deforms the pilot membrane, resulting in a pressure drop across the pilot stage which drives the opening of the main valve. The pilot pressure variations and the spring stiffness contribute to the actuation control in order to regulate the flow rate accurately and deliver a predefined fluid throughput as required by the user.

\begin{figure}
\centering
\includegraphics[width=0.6\textwidth]{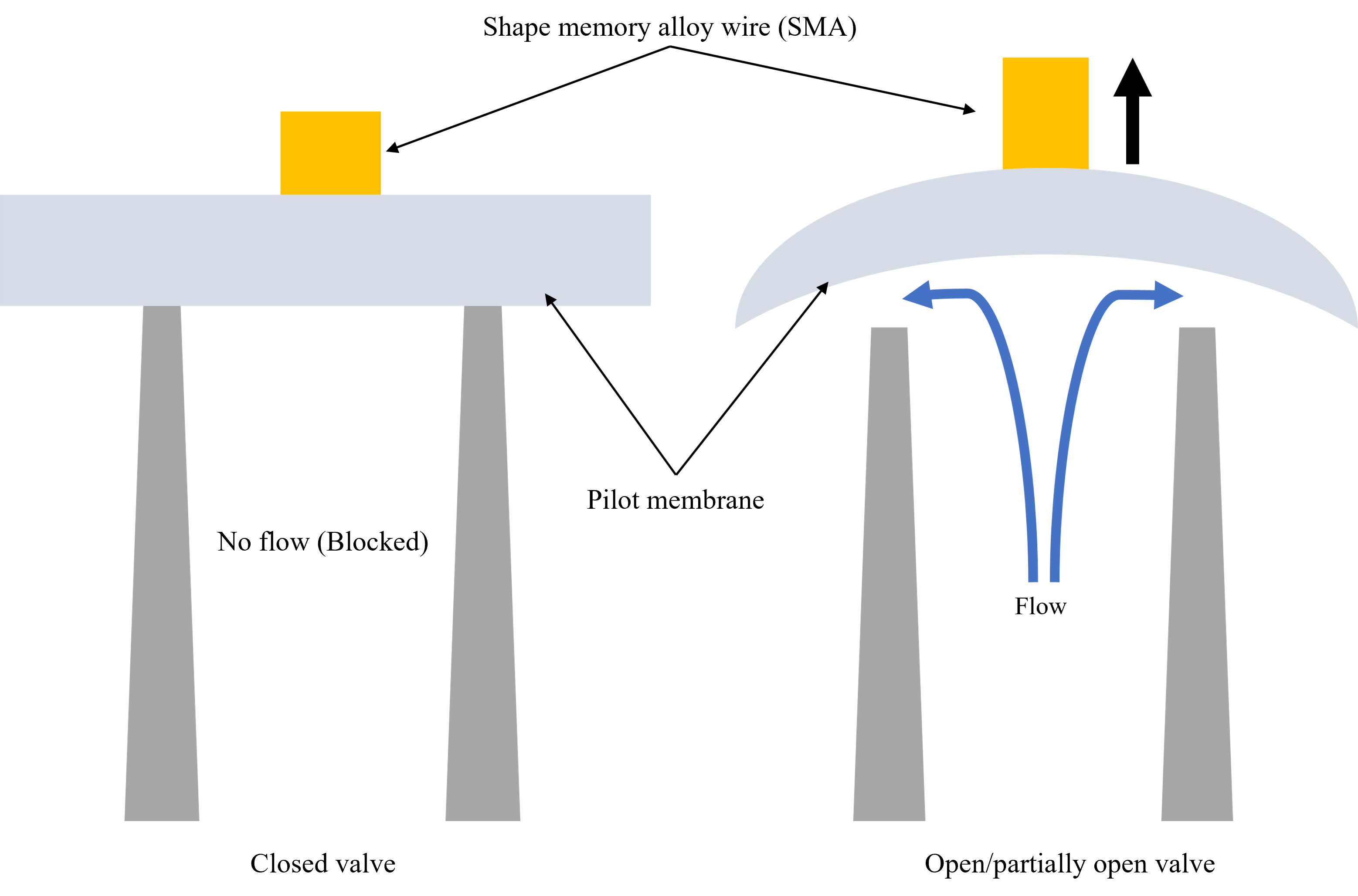}
\caption{Illustration of the actuation control based on shape memory alloy material (Yellow). The left-hand side corresponds to a closed POCV; the pilot membrane (Silver) is undeformed, and the upstream flow is blocked. The right-hand side corresponds to an actuated valve; the pilot membrane deforms according to the current SMA position, and the POCV switches to a partially or fully open position.}
\label{fig:actuations}
\end{figure}  

\begin{figure}
     \centering
     \includegraphics[width=0.8\textwidth]{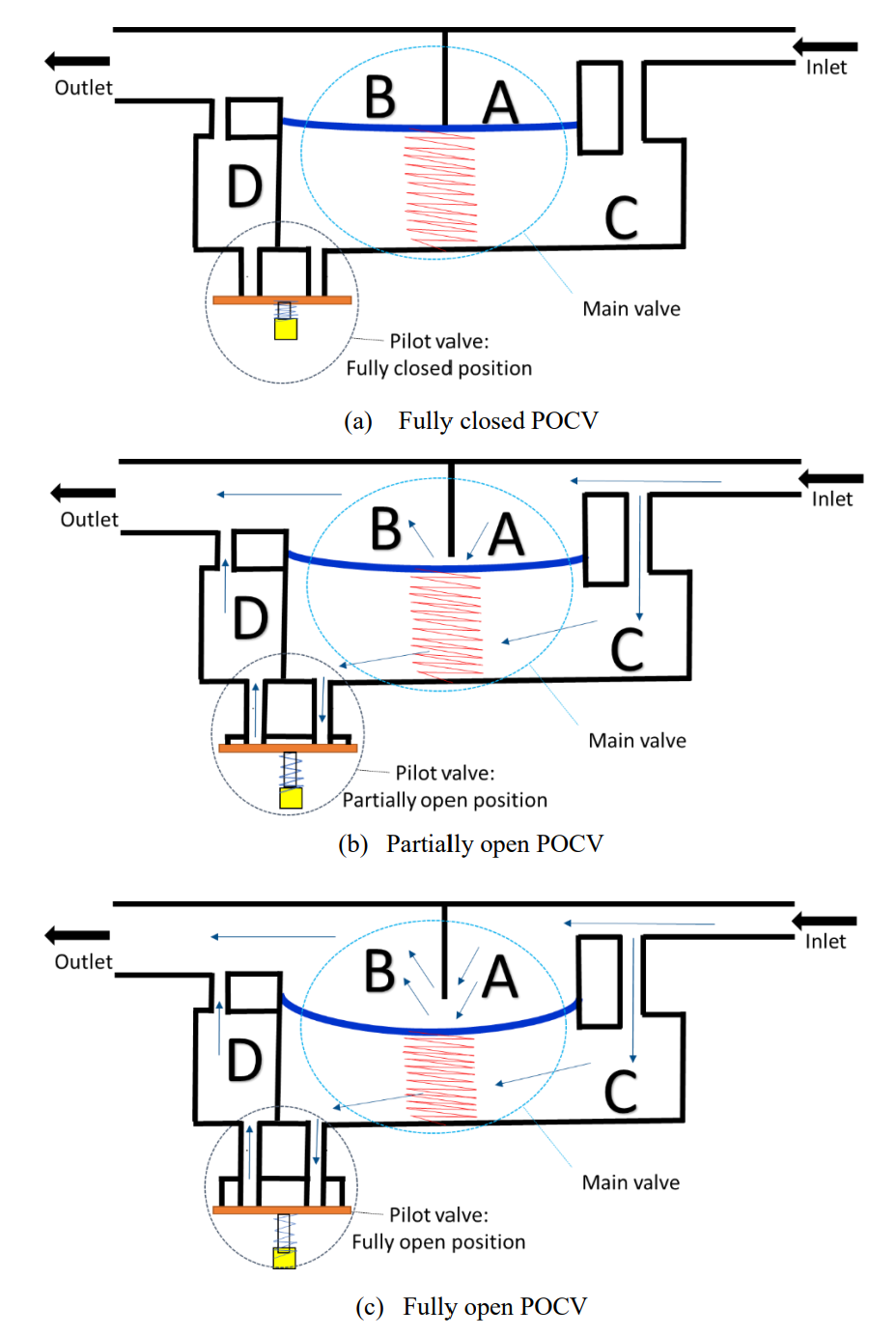}
        \caption{Schematic views of the pilot-operated valve under investigation for different opening setpoints. The valve can be operated as ON/OFF valve ((a) $\rightarrow$ (c)), and is designed to work as a proportional valve ((a) $\rightarrow$ (b) $\rightarrow$ (c)).  A: Inflow region; B: Outflow region; C: Pilot chamber region; D: Pilot chamber outflow.}
        \label{fig:three graphs}
\end{figure}

To elucidate the SMA-based actuation control mechanism, Figure~\ref{fig:actuations} shows the positions of the pilot membrane with respect to the SMA actuator: If the actuator is inactive, the POCV is in a normally closed mode (NC) ~\cite{Oh2006}. Otherwise, the updated SMA shape drives the pilot membrane position to allow fluid flow toward the outlet, which triggers the pressure change within the valve. In addition, given the reversible SMA shape effect, the valve can be opened and closed in a similar manner, and the two-way electro-thermal mechanical actuation enables multitudinous cycles during its lifespan. It is worth mentioning that the SMA-based actuation control yields remarkable benefits for fluidic applications, e.g. noiseless actuation, compact design, and lightweight products~\cite{Giulia_scalet}. The schematics in Figure~\ref{fig:three graphs} depict the general opening process of the valve in relation to the actuation control position; Panel (b), in which the main valve opens partially, pertains to the proportional actuation control for the POCV operation. \\
The POCV is designed to withstand high-pressure differences, ranging from $0.2$ up to $5$ bar while ensuring efficient sealing properties within the piping system. The technical information regarding dimensions and operating parameters pertinent to the present investigation, in accordance with the producer's datasheet, are listed in Table~\ref{tab:pocv}.

\begin{table}
\caption{Technical information relevant to the pilot-operated valve under study.}
\label{tab:pocv}
\centering
\begin{tabular}{cc}
\toprule
\textbf{Geometrical data}	& \\
\midrule
 Inlet orifice diameter [mm]	& $4$\\
 Outlet orifice diameter [mm]	& $8$\\
 Main membrane diameter [mm] &  $16.5$ \\
 Pilot membrane [mm] &         $2.7$\\
\toprule
\textbf{ Operating parameters}		& \\
\midrule
 Function  & Normally closed \\
 Activation & Pilot-operated \\
 Control & ON/OFF \\
 Pressure difference $\Delta P$ [bar] & $0.2$ $\cdot \cdot$  $5$\\
 Rise time [s] &  $0.15$ \\
 \bottomrule

\end{tabular}
\end{table}

\newpage


\section{Models and methods}
\label{sec:ModMet}

In this section, we present the fluid-structure interaction (FSI) model of the considered pilot-operated valve (POCV). In addition, we regard aspects of the numerical treatment of the FSI model.

\subsection{POCV model}
\begin{figure}
\centering
\includegraphics[width=0.8\textwidth]{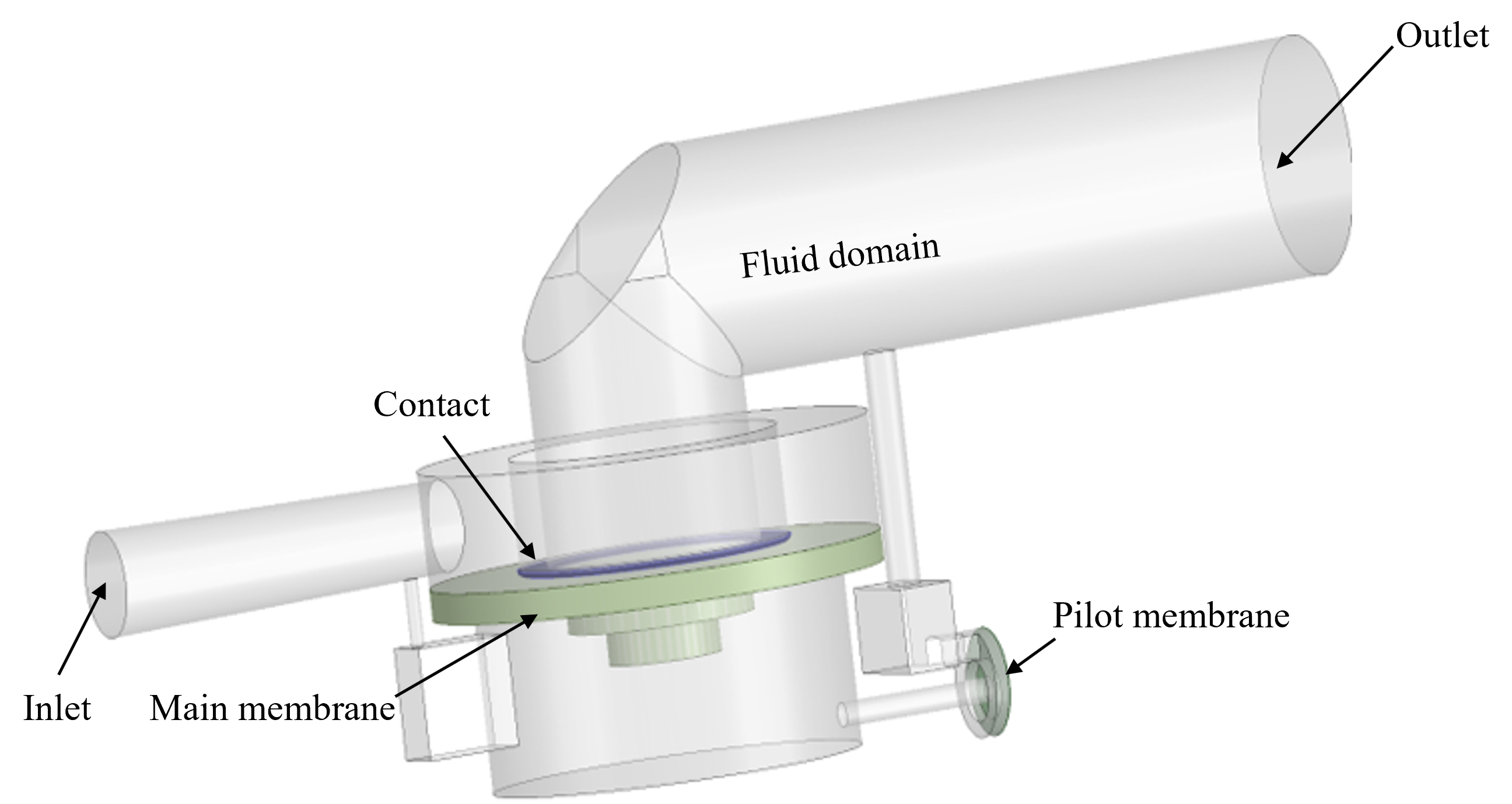}
\caption{Illustration of the pilot-operated control valve.}
\label{fig:CAD}
\end{figure}

The considered POCV configuration is illustrated in Figure~\ref{fig:CAD}. The POCV comprises a fluid domain and a structural domain and is surrounded by rigid exterior boundaries. The structural domain consists of two disjoint regions, viz., the pilot membrane and the main membrane. In the open configuration of the valve, the fluid domain is connected. In the closed configuration, the fluid domain is disconnected, comprising two components: one connected to the inlet of the valve, and one connected to the outlet. We denote by $\Omega\subset\mathbb{R}^3$ the POCV domain,
and by $\Omega_{\textsc{f}}(t)$ and $\Omega_{\textsc{s}}(t)$ the time-dependent domains occupied by the fluid and solid, respectively. The solid domain comprises two disjoint parts, $\Omega_{\textsc{s}}^{\textsc{m}}$ and~$\Omega_{\textsc{s}}^{\textsc{p}}$, corresponding to the main membrane and pilot membrane, respectively. The fluid-solid interface $\Gamma(t)$ corresponds to the intersection between the boundaries of the fluid and solid domains, according to $\Gamma(t)=\partial\Omega_{\textsc{f}}(t)\cap\partial\Omega_{\textsc{s}}(t)$. We denote by $\Gamma^{\textsc{m}}(t)$ (resp. $\Gamma^{\textsc{p}}(t)$) the fluid-solid interface associated with the main (resp. pilot) membrane.

\subsection{Fluid subsystem}
We consider a time interval $(0,T)$. Let 
$\boldsymbol{u}:(0,T)\times\Omega_{\textsc{f}}(t)\to\mathbb{R}^3$
and
$p:(0,T)\times\Omega_{\textsc{f}}(t)\to\mathbb{R}$ denote the fluid velocity and pressure, respectively. The fluid flow is described by the incompressible Navier--Stokes equations: 
\begin{subequations}
\label{eq:NS}
\begin{alignat}{2}
\rho_{\textsc{f}}\partial_t \boldsymbol{u}
+ \rho_{\textsc{f}} \boldsymbol{u} \cdot \nabla \boldsymbol{u} +\nabla p- \nabla\cdot\boldsymbol{\tau} &= 0 &\qquad &\text{in }(0,T)\times\Omega_{\textsc{f}}(t)
\label{eq:NS1}
\\
\nabla \cdot \boldsymbol{u} &=0 &\qquad &\text{in }(0,T)\times\Omega_{\textsc{f}}(t)
\end{alignat}
\end{subequations}
with $\rho_{\textsc{f}}$ as the fluid density. 
For closed or nearly closed valve configurations, the fluid velocities are sufficiently small to consider the flow as laminar. In this scenario, the stress tensor $\boldsymbol{\tau}$ coincides with the viscous stress and $\boldsymbol{\tau}=\boldsymbol{\tau}_{\textsc{v}}:=\mu_{\textsc{d}}(\nabla\boldsymbol{u}+(\nabla\boldsymbol{u})^T)$ with 
$\mu_{\textsc{d}}$ the dynamic viscosity of the considered fluid. For open configurations of the valve, the fluid tends to accelerate considerably due to the high upstream pressure and the flow enters the turbulent regime, viz., a regime in which the fluid exhibits flow separation and recirculation regions. To account for turbulent effects in the open configuration, we invoke the URANS model with Boussinesq closure. The stress tensor~$\boldsymbol{\tau}$ in~(\ref{eq:NS1}) is accordingly decomposed into a viscous part and a turbulent part as $\boldsymbol{\tau}=\boldsymbol{\tau}_{\textsc{v}}+\boldsymbol{\tau}_{\textsc{t}}$ where
\begin{equation}
\boldsymbol{\tau}_{\textsc{t}}=
\mu_{\textsc{t}}(\nabla\boldsymbol{u}+\big(\nabla\boldsymbol{u})^T\big)-\tfrac{2}{3}\rho_{\textsc{f}}k\boldsymbol{I}
\end{equation}
with $\mu_{\textsc{t}}$ the eddy viscosity, $k$ the turbulent kinetic energy and $\boldsymbol{I}$ the identity tensor. The eddy viscosity is generally closed in terms of the (averaged) fluid velocity by means of auxiliary variables. In particular, we consider the $k$\nobreakdash-$\omega$ shear stress transport (SST) turbulence model, where $\omega$ represents the specific dissipation rate~\cite{Menter1993}. In the $k$\nobreakdash-$\omega$ SST model, the eddy viscosity is given by
\begin{equation}
\label{eq:mut}
\mu_{\textsc{t}} = \frac{\rho_{\textsc{f}} a k}{{\rm max} (a \omega, \Omega F)} 
\end{equation}
where $a$ and $\Omega$ are model constants that depend on the specific implementation of the SST model, and $F:=F(k,\omega)$ is a specific function to transition between the $k$-$\omega$  model and the $k$-$\varepsilon$ model~\cite{menter1994two}. The SST turbulence model extends the fluid model by the aforementioned two auxiliary variables~$k$ and~$\omega$, subject to:
\begin{subequations}
\label{eq:SST}  
\begin{alignat}{2}
    \rho_{\textsc{f}} \partial_t k
    + \rho_{\textsc{f}} \nabla\cdot{}k \boldsymbol{u} 
    -\nabla\cdot(\Gamma_k\nabla{}k)&=  G_k - Y_k&\qquad &\text{in }(0,T)\times\Omega_{\textsc{f}}(t) \\
    \rho_{\textsc{f}} \partial_t \omega+ 
    \rho_{\textsc{f}} \nabla\cdot{}\omega\boldsymbol{u}-\nabla\cdot(\Gamma_{\omega}\nabla\omega) &=   G_{\omega} - Y_{\omega}&\qquad &\text{in }(0,T)\times\Omega_{\textsc{f}}(t) 
\end{alignat}
\end{subequations}
where $\Gamma_k$ and $\Gamma_{\omega}$ represent the effective diffusivities of $k$ and $\omega$, respectively. The nonnegative terms $G_k$ and $G_{\omega}$ (resp. $Y_k$ and $Y_{\omega}$) represent certain production (resp. dissipation) terms of turbulent kinetic energy and specific dissipation energy, respectively. We refer to~\cite{menter1994two} for further details and interpretation of the equations.

To accommodate the motion of the fluid domain in the fluid-structure interaction problem, the equation of motion~(\ref{eq:NS1}) and the turbulent balance laws~(\ref{eq:SST}) are reformulated in arbitrary Lagrangian-Eulerian (ALE) form; see, for instance, \cite{van2010fundamentals,Donea1982,farhat2004cfd,Bazilevs:2008kk}. This part of the formulation is standard in FSI and will not be further elaborated here.

The differential equations~(\ref{eq:NS}) and~(\ref{eq:SST}) must be furnished with suitable initial and boundary conditions. The boundary of the fluid domain comprises the fluid-solid interface at the main membrane, the wetted surface of the pilot membrane, rigid walls, an inflow boundary and an outflow boundary; see Figure~\ref{fig:CAD}. The boundary conditions at the fluid-solid interface are considered separately in Section~\ref{sec:IntCons}. At rigid solid walls, we impose the usual no-slip condition:
\begin{equation}
\label{eq:u=0}    
\boldsymbol{u}=0
\qquad\text{ at }(0,T)\times\Gamma_{\text{wall}}
\end{equation}
At the outlet, we impose a homogeneous traction condition according to:
\begin{equation}
\label{eq:BCout}
p\boldsymbol{n}-\boldsymbol{\tau{}n}=0
\qquad\text{ at }(0,T)\times\Gamma_{\text{out}}
\end{equation}
where $\boldsymbol{n}$ denotes the exterior unit normal vector on the boundary of the fluid domain.
It is to be noted that stability of the initial-boundary-value problem requires that $\boldsymbol{u}\cdot\boldsymbol{n}\geq{}0$ during the entire evolution of the flow field at parts of the boundary where~\EQ{BCout} is imposed.  At the inlet, we impose pressure in accordance with the pressure jump across the valve. Formally, this corresponds to a relation of the form~\EQ{BCout} with a non-zero right-hand side. However, as $\boldsymbol{u}\cdot\boldsymbol{n}$ is generally negative at the inflow boundary, the aforementioned stability condition associated with~\EQ{BCout} is violated at the inflow. Therefore, instead, at the inflow a scaled velocity profile is imposed:
\begin{equation}
\label{eq:BCin}
\boldsymbol{u}(t,\boldsymbol{x})=U(t)\,\hat{\boldsymbol{u}}(\boldsymbol{x})
\qquad\text{ at }(0,T)\times\Gamma_{\text{in}}
\end{equation}
where $\hat{\boldsymbol{u}}$ represents the prescribed inflow profile, compatible with the no-slip condition at the intersections of $\Gamma_{\text{in}}$ with the solid wall boundary $\Gamma_{\text{wall}}$. The time-dependent scaling factor $U$ is such that the average normal component of the traction at the inflow coincides with the prescribed inlet pressure~$p_{\text{in}}$:
\begin{equation}
\label{eq:pin}
\frac{1}{\operatorname{mean}(\Gamma_{\text{in}})}
\int_{\Gamma{\text{in}}}
p-\boldsymbol{n}\cdot\boldsymbol{\tau{}n}=p_{\text{in}}  
\end{equation}
At the section $\Gamma^{\textsc{p}}(t)$ of the boundary coinciding with the wetted boundary of the pilot membrane (see Figure~\ref{fig:CAD}), the fluid domain is deformed in accordance with the deformation of the pilot membrane, and the fluid velocity coincides with the velocity of the pilot membrane:
\begin{equation}
\label{eq:pilotu}
\boldsymbol{u}=\partial_t\boldsymbol{d}_{\textsc{s}}\qquad\text{at }(0,T)\times\Gamma^{\textsc{p}}(t)
\end{equation}
where $\boldsymbol{d}_{\textsc{s}}$ stands for the deformation of the solid, in particular, the pilot membrane. Boundary condition~\EQ{pilotu} can be conceived of as so-called `one-way FSI', in that the fluid domain deforms in accordance with the deformation of the solid, and the fluid velocity coincides with the solid velocity, but the solid does not respond to the traction exerted by the fluid on the wetted boundary of the pilot membrane; see also Section~\ref{sec:IntCons}.

A detailed treatment of boundary conditions for the SST turbulence model~\EQ{SST} is beyond the scope of this work. At solid walls, including the pilot membrane, homogeneous and non-homogeneous Dirichlet-type boundary conditions are typically imposed on $k$ and $\omega$, respectively:
\begin{subequations}
    \begin{alignat}{2}
        k&=0\qquad &\text{at }(0,T)\times(\Gamma_{\text{wall}}\cup\Gamma^{\textsc{p}})
        \\
        \omega&=\omega_{\text{wall}}\qquad &\text{at }(0,T)\times(\Gamma_{\text{wall}}\cup\Gamma^{\textsc{p}})
    \end{alignat}
\end{subequations}
where $\omega_{\text{wall}}$ represents an estimate of the dissipation rate at the boundary. At the outflow boundary, homogeneous Neumann conditions are imposed:
\begin{subequations}
    \begin{alignat}{2}
        \boldsymbol{n}\cdot\nabla{}k&=0\qquad &\text{at }(0,T)\times\Gamma_{\text{out}}
        \\
        \boldsymbol{n}\cdot\nabla{}\omega&=0\qquad &\text{at }(0,T)\times\Gamma_{\text{out}}
    \end{alignat}
\end{subequations}
At the inflow part of the boundary, one generally imposes non-homogeneous Dirichlet conditions on~$k,\omega$, according to
\begin{subequations}
    \begin{alignat}{2}
        k&=k_{\text{in}}\qquad &\text{at }(0,T)\times\Gamma_{\text{in}}
        \\
        \omega&=\omega_{\text{in}}\qquad &\text{at }(0,T)\times\Gamma_{\text{in}}
    \end{alignat}
\end{subequations}
where $k_{\text{in}}$ and~$\omega_{\text{in}}$ represent exogenous data, corresponding to estimates of the turbulent kinetic energy and specific dissipation rate at the inlet.

The initial conditions for~(\ref{eq:NS}) and~(\ref{eq:SST}) comprise a specification of~$\boldsymbol{u},k,\omega$ at~$t=0$. In accordance with the fact that the valve is initially in a closed configuration, we impose a homogeneous initial condition on the velocity:
\begin{equation}
    \boldsymbol{u}(t=0)=0\qquad\text{in }\Omega_{\textsc{f}}(t=0)
\end{equation}
Because the fluid is stagnant in the initial closed configuration, the turbulent kinetic energy and specific dissipation rate vanish initially:
\begin{subequations}
    \begin{alignat}{2}
        k(t=0)&=0\qquad &\text{in }\Omega_{\textsc{f}}(t=0)
        \\
        \omega(t=0)&=0\qquad &\text{in }\Omega_{\textsc{f}}(t=0)
    \end{alignat}
\end{subequations}


\subsection{Structure subsystem}
\label{sec:struct}
\begin{figure}
\centering
\includegraphics[scale=0.4]{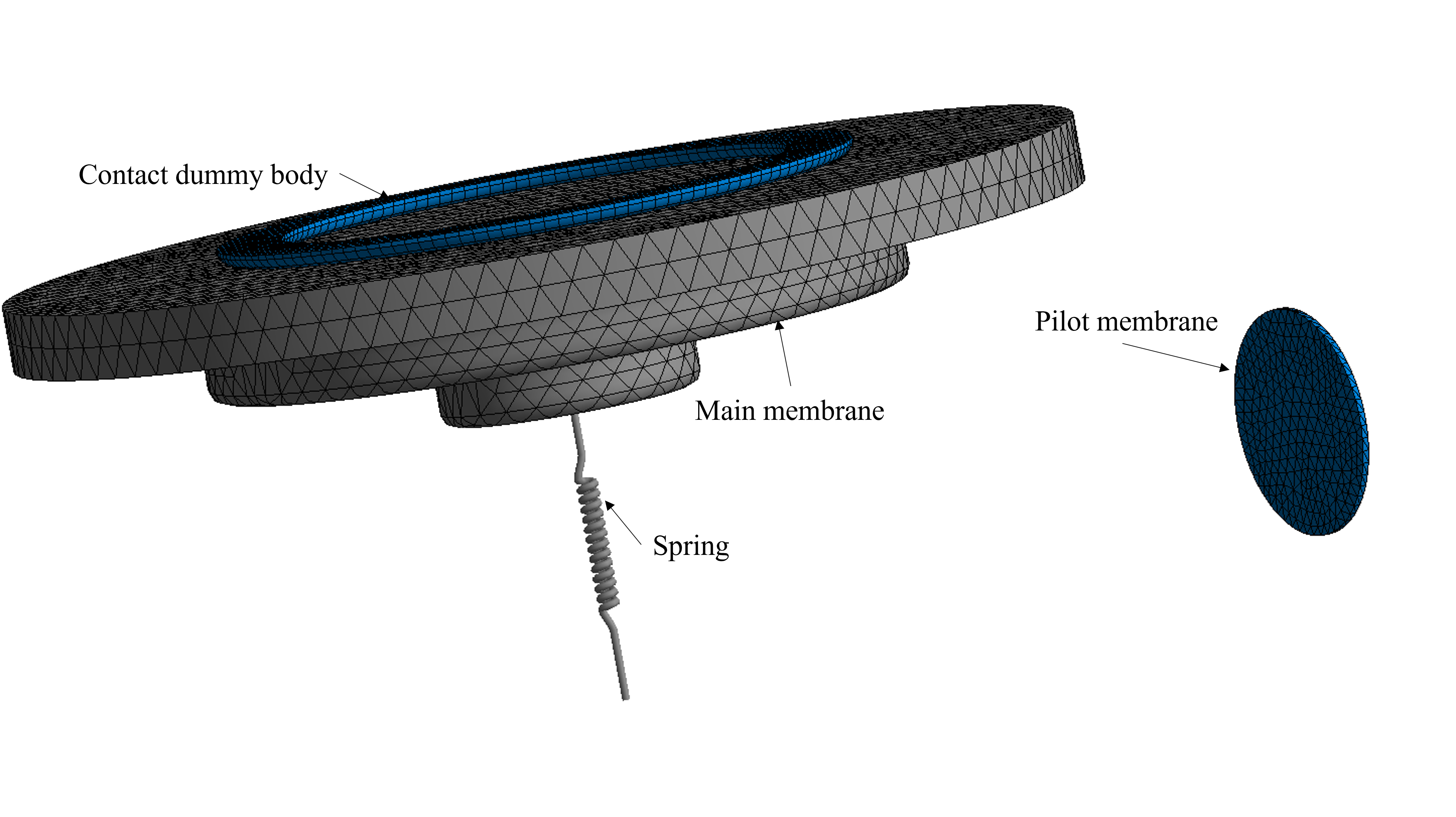}
\caption{POCV structural system after spatial discretization. The main membrane is constrained by a contact dummy body representing the valve seat and a spring to guide the deformation. The pilot membrane is clamped and controlled by a displacement boundary condition to replicate the SMA-based actuator actions.}
\label{fig:strut}
\end{figure}

The structure subsystem is illustrated in 
Figure~\ref{fig:strut}. The structure comprises three components, viz. the main membrane, the pilot membrane, and a quasi-rigid contact dummy body. The dummy body represents the so-called valve seat, which is the part of the casing of the POCV which comes into contact with the main membrane. The dummy body is further elaborated in Section~\ref{sec:NumApp}. The pilot membrane is modelled as an independent actuator, in the sense that a prescribed motion is imposed on part of its external (non-wetted) boundary, representing the actuation, and the pilot membrane is ignorant to the presence of the fluid at its wetted boundary. The main membrane, on the other hand, does respond to the fluid-traction exerted on its wetted boundary. The main membrane, therefore, represents the only deformable structure component in the aggregated fluid-structure interaction problem. For the structural subsystem, however, this essential difference between the main membrane and the pilot membrane only manifest as a difference in the load. Hence, we present the model for the structure subsystem, comprising both main and pilot membrane, in a unified manner.

The membranes are composed of an incompressible rubber-like material. 
The time-dependent deformation of the structural domain is rendered in an updated-Lagrangian formulation in the numerical approximation. However, to facilitate and condense the presentation, we present the model in the equivalent total-Lagrangian formulation. The membranes exhibit dissipation under dynamic deformation.
The dissipation is modelled by Rayleigh damping. Because Rayleigh damping is most conveniently expressed in a weak formulation, we regard the structural model in weak form.
Denoting by $\hat{\Omega}_{\textsc{s}}=\Omega_{\textsc{s}}(t=0)$ the reference configuration of the solid subsystem,
the deformation $\boldsymbol{d}_{\textsc{s}}:(0,T)\times\hat{\Omega}_{\textsc{s}}\to\Omega_{\textsc{s}}(t)$ satisfies:
\begin{subequations}
\label{eq:solid}
\begin{align}
\int_{\hat{\Omega}_{\textsc{s}}}
\rho_{\textsc{s}} 
\boldsymbol{w}\cdot\frac{\partial^2 \boldsymbol{d}_{\textsc{s}}}{\partial t^2}
+
\int_{\hat{\Omega}_{\textsc{s}}}
\operatorname{Grad}\boldsymbol{w}:\boldsymbol{P}
+
\int_{\hat{\Omega}_{\textsc{s}}}
\alpha\rho_{\textsc{s}} 
\boldsymbol{w}\cdot\frac{\partial \boldsymbol{d}_{\textsc{s}}}{\partial t}
\quad
&
\nonumber
\\
+
\int_{\hat{\Omega}_{\textsc{s}}}
\beta
\operatorname{Grad}\boldsymbol{w}:\boldsymbol{P}'(\,\,\hat{\!\!\boldsymbol{d}}_{\textsc{s}},\partial_t\boldsymbol{d}_{\textsc{s}})
&=l(\boldsymbol{w})\phantom{0}
\quad 
\text{in}\ (0,T) \times \hat{\Omega}_{\textsc{s}}
\label{eq:solid1}
\\
\int_{\hat{\Omega}_{\textsc{s}}}
q(\det{}\boldsymbol{F}-1)&=0\phantom{l(\boldsymbol{w})}
\quad 
\text{in}\ (0,T) \times \hat{\Omega}_{\textsc{s}}
\end{align}
\end{subequations}
for all admissible test functions $\boldsymbol{w},q$,
where $\rho_s$ is the solid mass density, $\boldsymbol{F}=\operatorname{Grad}\boldsymbol{d}_{\textsc{s}}$ denotes the deformation gradient, and $\boldsymbol{P}$ is the first Piola--Kirchhoff stress tensor. The load functional in the right-hand side of~\EQ{solid1} vanishes for the pilot membrane, and is further elaborated in Section~\ref{sec:IntCons} for the main membrane.
It is to be noted that the gradient operator in~\EQ{solid}, $\operatorname{Grad}$, acts in the reference configuration. The last two terms in the left-hand side of~\EQ{solid1} represent Rayleigh damping,
where $\boldsymbol{P}'(\,\,\hat{\!\!\boldsymbol{d}}_{\textsc{s}};\partial_t\boldsymbol{d})$ denotes the directional derivative of $\boldsymbol{P}(\boldsymbol{d})$ with respect to~$\boldsymbol{d}$, evaluated in the reference configuration~$\,\,\hat{\!\!\boldsymbol{d}}_{\textsc{s}}$ ,
 in the direction $\partial_t\boldsymbol{d}$. One may note that~$\,\,\hat{\!\!\boldsymbol{d}}_{\textsc{s}}(\boldsymbol{X})=\boldsymbol{X}$. The coefficients $\alpha,\beta\geq{}0$ are referred to as the Rayleigh coefficients.
The damping term is such that in a finite-element approximation of~\EQ{solid}, it leads to a term $\mathbf{D}\,d_t\mathbf{d}$ with $\mathbf{D}=\alpha\mathbf{M}+\beta\mathbf{T}_0$, 
where~$\mathbf{M}$ and~$\mathbf{T}_0$ denote the mass matrix and tangent stiffness matrix at~$t=0$, respectively.

A suitable class of constitutive models to describe the elastic behavior of the considered rubber-like material is provided by the hyperelastic material models. Correspondingly, the first Piola--Kirchhoff stress tensor is the  derivative of an associated stored-energy density function $W:=W(\boldsymbol{F})$ with respect to the deformation gradient. Denoting the Green--Lagrange strain tensor by $\boldsymbol{E}:=\boldsymbol{E}(\boldsymbol{F})=\frac{1}{2}(\boldsymbol{F}^T\boldsymbol{F}-\boldsymbol{I})$, where $\boldsymbol{I}$ is the identity tensor, the material of the membranes is characterized by a neo-Hookean constitutive relation, associated with the stored-energy function:
\begin{equation}
\label{eq:neoHookean}
     W(\boldsymbol{F}) = C_{10}\operatorname{tr}(2\boldsymbol{E}_{\textsc{s}})-p_{\textsc{s}}\big(\sqrt{\det{}(2\boldsymbol{E}+\boldsymbol{I})}-1\big)
\end{equation}
where $C_{10}$ is a material constant and $p_{\textsc{s}}$ is  pressure. It is to be noted that $\operatorname{tr}(2\boldsymbol{E})=I_1-3$ with $I_1$ the first strain invariant and $\sqrt{\det(2\boldsymbol{E}+\boldsymbol{I})}=\det{}\boldsymbol{F}$. The material parameter~$C_{10}$, which characterizes the material response, is calibrated on the basis of experimental data; see~\ref{calib}.

In the POCV device, a spring is mounted to the main membrane to provide an additional displacement-proportional load; see Figure~\ref{fig:strut}. The spring is modelled in the structure subsystem by a force 
$-k(\boldsymbol{d}_{\textsc{s}}(t,\hat{\boldsymbol{X}}_{\textsc{spr}})-\hat{\boldsymbol{X}}_{\textsc{spr}})$, where $k$ indicates the spring constant, and $\hat{\boldsymbol{X}}_{\textsc{spr}}$ is the mount point of the spring in the reference configuration. As elastic solids theoretically do not admit point loads, the spring force is distributed on a region of the membrane surface, $\Gamma_{\textsc{spr}}(t)\subset\partial\Omega_{\textsc{s}}(t)$. 
The spring is hence represented by a traction:
\begin{equation}
\label{eq:spring}
\boldsymbol{t}_{\textsc{spr}}(\boldsymbol{x})
=
\begin{cases}
-k(\boldsymbol{d}_{\textsc{s}}(t,\hat{\boldsymbol{X}}_{\textsc{spr}})-\hat{\boldsymbol{X}}_{\textsc{spr}})/\operatorname{Area}(\Gamma_{\textsc{spr}})
&\text{if }\boldsymbol{x}\in\Gamma_{\textsc{spr}}
\\
0&\text{otherwise}
\end{cases}
\end{equation}
acting on the surface of the membrane.

The differential equations~\EQ{solid} must be provided with suitable initial and boundary conditions. The boundary-conditions at the wetted boundary $\Gamma^{\textsc{m}}(t)\subset\partial\Omega_{\textsc{s}}(t)$ of the main membrane, corresponding to the fluid-solid interface, represented by the load functional in the right member of~\EQ{solid1}, are presented in Section~\ref{sec:IntCons}. Let us note that the aforementioned spring load is administered at a part of the wetted boundary, i.e., $\Gamma_{\textsc{spr}}(t)\subseteq\Gamma^{\textsc{m}}(t)$.
On parts of the boundary where the membranes are fixed to the casing of the POCV, $\Gamma_{\textsc{fix}}$, the structural configuration is identified with the initial configuration, i.e.,
\begin{equation}
\boldsymbol{d}_{\textsc{s}}(t,\boldsymbol{X})
=\boldsymbol{X}
\quad\text{for all }(t,\boldsymbol{X})\in(0,T)\times\Gamma_{\textsc{fix}}
\end{equation}
For the main membrane, the fixed boundary and wetted boundary are complementary. For the pilot membrane, the remaining part of the boundary, including the wetted boundary, is subject to homogeneous traction (i.e., stress-free) boundary conditions, except for a small region at the center of the external (non-wetted) boundary, where the deformation in accordance with the actuation is imposed.

Suitable initial conditions for~\EQ{solid} are provided by a specification of the initial deformation, $\boldsymbol{d}_{\textsc{s}}(t=0)$, and the initial velocity, $\partial_t\boldsymbol{d}_{\textsc{s}}(t=0)$. Noting that the configuration of the membrane at $t=0$ coincides with the initial configuration, it holds that:
\begin{equation}
\boldsymbol{d}_{\textsc{s}}(t=0,\boldsymbol{X})=\boldsymbol{X}
\quad\text{for all }\boldsymbol{X}\in\Omega_{\textsc{s}}(t=0)
\end{equation}
We assume that both membranes are initially at rest and, hence,
\begin{equation}
\partial_t\boldsymbol{d}_{\textsc{s}}(t=0)=0
\quad\text{in }\Omega_{\textsc{s}}(t=0)
\end{equation}

\subsection{Contact formulation}
\label{sec:contact}
The deformation of the main membrane is confined by the rigid parts of the POCV, notably the so-called valve seat. In the closed configuration of the POCV, the main membrane is in contact with the valve seat in such a manner that the fluid domain is separated into two disjoint parts. In the fully open configuration, there is no contact between the membrane and the valve seat. The deformation of the elastic main membrane is subject to a non-penetration condition of the form 
\begin{equation}
\label{eq:nonpen}
\operatorname{dist}(\partial\Omega_{\textsc{s}}(t),\Gamma_{\textsc{vs}})\geq{}0,     
\end{equation}
where $\Gamma_{\textsc{vs}}\subset\partial\Omega_{\textsc{f}}(t=0)$ denotes the part of the rigid boundary of the POCV corresponding to the valve seat. In the fully open configuration, the inequality in~\EQ{nonpen} is strict, i.e., the distance between the membrane boundary and the valve seat is strictly positive. Contact occurs if equality holds in~\EQ{nonpen}.
An extensive review of contact mechanics is beyond the scope of the present work: the reader is referred to~\cite{yastrebov2013numerical} and~\cite{popp2012mortar} for a review of existing contact formulations.

\begin{figure}
\centering
\includegraphics[scale=0.5]{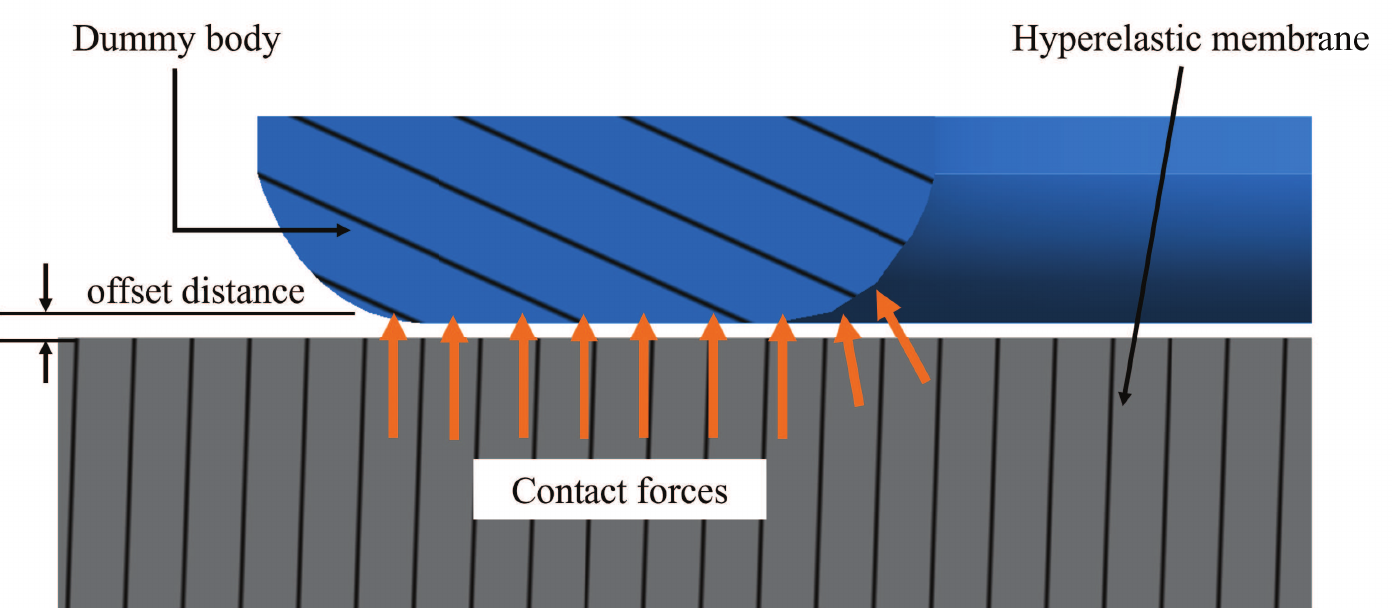}
\caption{Membrane to valve seat inter-body contact modelling. Blue: Contact dummy body section. Grey: Main membrane section. The mechanical parts are separated by a thin gap of thickness $\epsilon>0$ equal to the fluid gap width for consistency.}
\label{fig:contact}
\end{figure}

A fundamental complication in numerical procedures for fluid-solid-contact interaction problems~\cite{Ager:2021rf}, pertains to the fact that the topological change in the fluid domain occurring at contact (from a simply connected set to connected set with holes [partial contact] or a disconnected set [fully closed]) manifests in the computational method as a collapse of the mesh in the fluid domain. To avoid such a collapse of the mesh, it is generally necessary to ensure a small-but-finite separation of the surfaces, i.e., that the inequality in~\EQ{nonpen} holds strictly. To this purpose, we modify the contact formulation in a way that it retains (aims to retain)  a thin gap of thickness $\epsilon>0$ between the valve seat and the main membrane; see Figures~\ref{fig:strut} and~\ref{fig:contact}. This implies that the contact load is activated if the distance between the main membrane and valve seat is less than or equal to~$\epsilon$ 
thus ensuring a small-but-finite separation between the main membrane and the actual valve seat at contact. Simultaneously, the fluid domain in the contact region is replaced by a Darcy-type flow resistance if contact is detected, to mimic the sharp increase in the flow resistance that occurs as the separation between the valve and the valve seat approaches zero.

Contact is modelled by means of an auxiliary contact traction on the elastic structure. The valve seat is represented by a quasi-rigid dummy body; see Figures~\ref{fig:strut} and~\ref{fig:contact}.
Noting that the fluid between the two contacting surfaces effectively lubricates the contact, we ignore friction and, hence, the contact force acts in the direction perpendicular to the contact surface. The considered contact traction (force per unit area) is of the form:
\begin{equation}
\label{eq:cforce}
\boldsymbol{t}_{\textsc{c}}(\boldsymbol{x})
=
-\boldsymbol{n}_{\textsc{s}}
\big(
k_{\textsc{c}}
\big\lfloor(\boldsymbol{x}-\mathcal{P}\boldsymbol{x})\cdot\boldsymbol{n}_{\textsc{s}}+\epsilon\big\rfloor+\lambda_{\textsc{c}}\big)
\end{equation}
where $\mathcal{P}\boldsymbol{x}$ denotes the projection of the point $\boldsymbol{x}$ on the surface of the membrane~$\partial\Omega_{\textsc{s}}(t)$ (resp. the surface of the dummy body $\Gamma_{\textsc{d}}(t)$) along the normal vector onto~$\Gamma_{\textsc{d}}(t)$ (resp.~$\partial\Omega_{\textsc{s}}(t)$), the operator $\lfloor\cdot\rfloor=\max(\cdot,0)$ is the positive part (or Macaulay bracket) and $\lambda_{\textsc{c}}$ is the contact force, i.e., the Lagrange multiplier associated with the non-penetration condition.
The first term in parentheses in~\EQ{cforce} represents a penalty contribution, the penalty parameter $k_{\textsc{c}}>0$ being analogous to an elastic-spring constant. By virtue of the Macaulay bracket and the offset parameter~$\epsilon$, the contact traction is active only in regions where the signed distance between the membrane and the dummy body is less than~$\epsilon$.

\subsection{Interface conditions of the FSI problem}
\label{sec:IntCons}
The fluid and the structure are interconnected by kinematic and dynamic interface conditions~\cite{van2010fundamentals}. The kinematic conditions prescribe that the boundaries of the fluid and solid domains coincide at the fluid-solid interface, and that the fluid velocity coincides with the structure velocity at the interface. The latter condition implies,
\begin{equation}
\label{eq:kinematic}
\boldsymbol{u}=\partial_t\boldsymbol{d}_{\textsc{s}}
\quad\text{in }(0,T)\times\Gamma(t)
\end{equation}
Let us note that on the left-hand (resp. right-hand) side  of~\EQ{kinematic}, positions are to be understood in the actual (resp. reference) configuration. The kinematic condition ~\EQ{kinematic} can be conceived of as an essential (Dirichlet-type) boundary condition for the fluid subsystem.


The dynamic condition enforces the balance of tractions at the fluid-solid interface. For the main membrane, the load functional in~\EQ{solid1} therefore comprises the fluid traction, in addition to the traction exerted by the spring and by contact, according to~\EQ{spring} and~\EQ{cforce}, respectively:
\begin{equation}
\label{eq:load}
l(\boldsymbol{w})
=
\int_{\Gamma(t)}
\boldsymbol{w}\cdot(
p\boldsymbol{n}_{\textsc{f}}-\boldsymbol{\tau}\boldsymbol{n}_{\textsc{f}})
+
\int_{\Gamma(t)}
\boldsymbol{w}\cdot\boldsymbol{t}_{\textsc{c}}
+
\int_{\hat{\Gamma}}
\boldsymbol{w}\cdot\boldsymbol{t}_{\textsc{spr}}
\end{equation}
It is to be noted that the first and second terms in~\EQ{load}, corresponding to the traction exerted by the fluid on the solid and the contact traction, are formulated in the actual configuration, while the third term, associated with the distributed spring load, is expressed in the reference configuration. The dynamic condition in~\EQ{load} can be regarded as a natural (Neumann-type) boundary condition for the structure subsystem.

\subsection{Numerical approximation of the fluid and solid subsystems}
\label{sec:NumApp}
The fluid calculations are executed using the commercial software package ANSYS Fluent and are based on an approximation of the URANS transport equations by means of a finite-volume method in an ALE formulation. A second-order upwind scheme is used for the spatial discretization of the convective terms and a least-squares 
cell-based formulation is used to approximate the diffusive fluxes in the momentum equation. 

\begin{figure}
\centering
\includegraphics[scale=0.5]{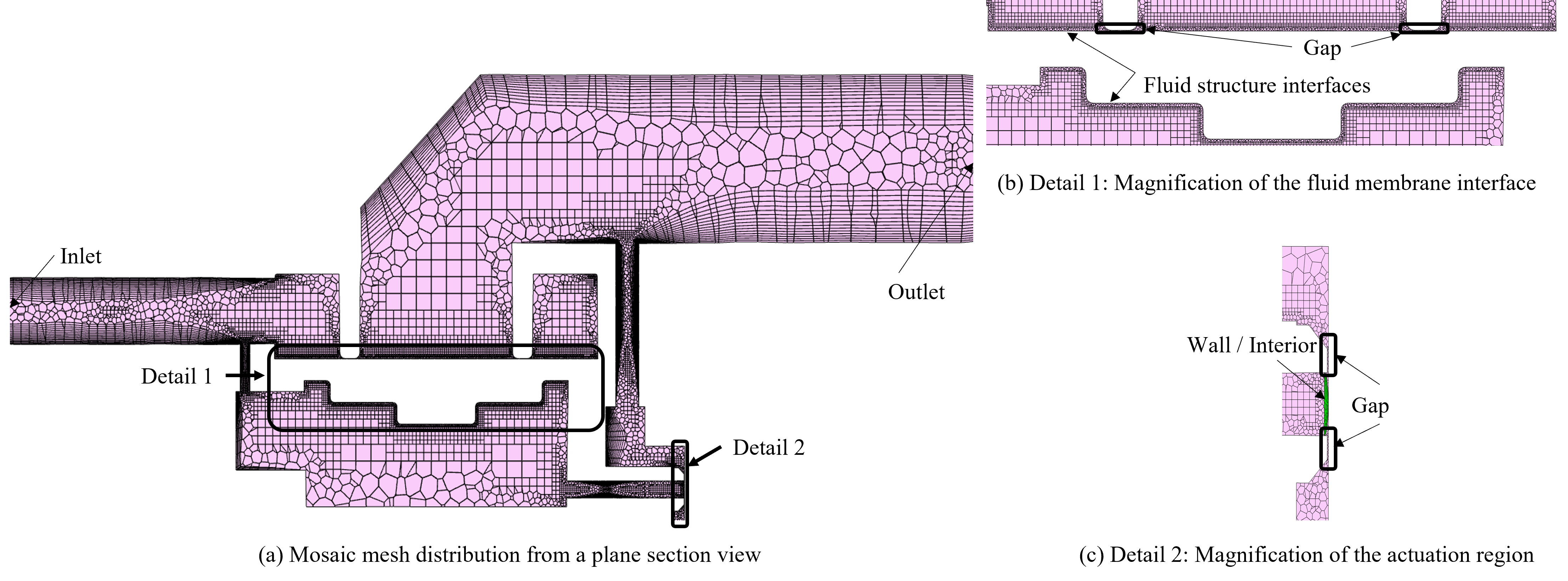}
\caption{Fluid mesh from a plane section view. The stationary (non-moving) zones are refined using inflation layers to improve accuracy. The deforming regions are refined using smaller cells to reduce the probability of mesh folding due to excessive motion. (b) Detail 1 shows the fluid-structure interfaces and the adjacent deforming zones. (c) Detail 2 depicts the pilot membrane and adjacent moving domain. Full closure of the valve is performed using a wall boundary condition (pilot region) and a gap model (main flow section).}
\label{fmesh}
\end{figure}

Figure~\ref{fmesh} illustrates the fluid-mesh partition from a plane-section-view perspective. The mesh combines different element types. The bulk is composed of octree hexagonal elements. Boundary-layer prisms are used in the vicinity of rigid boundaries. The regions adjacent to the moving surfaces are mostly of general polyhedral type. 
The region adjacent to the fluid-solid interface is refined to capture the large gradients in this area, as shown in panels~(b) and~(c) of Figure~\ref{fmesh}. This heterogeneous mesh layout serves to enhance the accuracy while restricting the number of degrees of freedom in the discrete approximation. The fluid mesh used in the numerical simulations comprises $1.019.544$ cells. 

The pilot channel is closed using a wall boundary zone, while the main membrane blocks the flow using a gap model. This gap model is active when the distance between the FSI wet interface and the valve seat is below a predefined contact distance threshold; see also 
section~\ref{sec:contact}. A spring-based smoothing method controls the deforming mesh dynamics. To improve the robustness of the mesh-deformation algorithm, 
a strict tolerance is used on the dynamic mesh convergence criterion, and under-relaxation of the mesh motion parameter is applied. 

The solution procedure for the discretized fluid equations is based on concurrent pressure-velocity coupling, instead of segregated approaches such as SIMPLE and PISO. The choice for a coupled solution procedure is motivated by its improved robustness relative to segregated solution approaches, which is particularly important in view of the moving mesh and the strongly varying flow velocity (encoded by the local Reynolds number) within the flow field, especially during the valve-closing event. In our experience, concurrent pressure-velocity coupling yields a better compromise between computational complexity, time-step-size restrictions, and robustness, compared to segregated solvers. To retain accuracy and robustness throughout the evolution of the valve configuration, it is important to ensure convergence of the CFD solver to a strict tolerance. In particular, in addition to the ANSYS Fluent default criterion that the scaled residual drops below~$10^{-3}$, we require that the relative pressure updates near the fluid-solid interface and near the pilot membrane drop below~ $10^{-4}$.

The structure is discretized using tetrahedral Taylor-Hood elements with quadratic displacements and linear pressures. The mesh partition of the main membrane is displayed in Figure~\ref{fig:strut}. The main-membrane domain is locally refined near the valve-seat contact region and contains $24.551$ elements. The 
pilot-membrane's mesh is uniformly distributed and comprises $1.555$ elements. 

Because of differences in their geometric representation, ANSYS Fluent does not admit to directly monitor contact between an endogenous structural element and an exogenous rigid surface, e.g. the main membrane and the casing of the POCV. To facilitate the contact formulation in ANSYS Fluent, the valve seat, which is the part of the casing of the POCV that comes into contact with the main membrane, is therefore represented as a separate quasi-rigid dummy body in the computational model.

The solution procedure for the structure subsystem, including contact, consists of a standard Newton procedure, augmented with a time-step-reduction procedure which acts as a fail-safe in case of divergence or excessively slow convergence of the Newton process. If time-step reduction is invoked, the structure subproblem is subcycled until it is synchronized with the fluid, i.e., multiple smaller time steps are carried out.
The linear tangent problems in the Newton procedure are solved by means of a sparse direct solver.

\subsection{FSI solution strategy}
\label{sec:FSIsol}
Approaches for solving fluid-structure interaction problems can generally be classified as monolithic or partitioned~\cite{farhat2004cfd,van2010fundamentals}. In the monolithic approach, the aggregated system of equations is solved in a fully coupled manner at each time step, i.e., the fluid and structure subsystems are solved simultaneously by means of one solver. The main advantage of monolithic methods is their robustness. However, a significant disadvantage is the inherent loss of modularity, obstructing the reuse of established solution strategies and software frameworks. In addition, monolithic methods generally carry a high computational complexity~\cite{van2010fundamentals}. 
Partitioned methods are iterative procedures, in which the fluid and structural subsystems are solved separately, and which (in principle) yield the solution of the coupled FSI problem upon convergence. The standard iterative procedure is referred to as {\em subiteration\/}, and consists in solving the fluid subsystem subject to the kinematic condition, and the structure subsystem subject to the dynamic condition. By virtue of the intrinsic partitioning, partitioned methods are inherently modular, and enable reuse of existing methodologies that have been developed for the fluid and solid subsystems. Consequently, partitioned methods are much more flexible and versatile than monolithic methods. A further subcategorization of partitioned methods can be considered, into loosely coupled and strongly coupled techniques. In loosely coupled (or {\em staggered\/}) approaches~\cite{farhat2004cfd,piperno}, the fluid and solid subsystems are solved only once per time step. In strongly coupled approaches~\cite{van2011partitioned}, the iteration is continued until a prescribed convergence criterion is satisfied. 

The coupling strength in FSI problems is generally characterized by the so-called {\em added mass\/}. This characterization is based on the notion that the force exerted by the fluid on the solid is essentially proportional to the acceleration of the solid, and the fluid therefore acts on the structure as an added mass~\cite{causin2005added,van2009added,Forster:2007aa}. In partitioned methods, the fluid and the solid are treated asynchronously and, hence, the added mass of the fluid is treated explicitly. Consequently, strongly-coupled FSI problems, i.e., FSI problems with large added-mass effects, form a vulnerability of partitioned methods, leading to divergence or prohibitively slow convergence. Various techniques have however been developed to stabilize and accelerate strongly-coupled partitioned methods based on subiteration, e.g, underrelaxation, Aitken's method, Quasi-Newton methods~\cite{DEGROOTE2009793,HAELTERMAN20169}, or (interface) artificial compressibility~\cite{degrooteAC,Bogaers2015}
; see~\cite{van2011partitioned} for an overview. It is to be noted that artificial-compressibility methods pertain specifically to incompressible FSI, based on the fact that the incompressibility of the fluid causes its added-mass to be essentially independent of the time step~\cite{van2009added}.

An additional problem pertaining to enclosed incompressible FSI problems, is the so-called {\em incompressibility dilemma\/}~\cite{Kuttler_incompress}. In enclosed incompressible FSI problems, the structure can only assume deformations that are compatible with the volume of the fluid. This compatibility condition~\cite{timo_van2015} causes severe instability in partitioned methods, if it is not properly handled. This instability is essentially independent of the added mass of the fluid, and it cannot be resolved by most of the stabilization techniques intended to address added-mass effects, such as under-relaxation, quasi-Newton methods, and Aitken's method. Artificial-compressibility stabilization is effective, as it replaces the incompressible FSI problems by a sequence of compressible ones.

The considered POCV problem represents a strongly coupled FSI problem which, in addition, displays characteristics of the incompressibility dilemma. The fluid carried by the POCV is typically aqueous, and is essentially incompressible. The mass density of the material of the main membrane is similar to that of the carried fluid. Hence, the POCV system displays a significant added-mass effect. Moreover, in the closed configuration of the POCV, the upstream fluid volume is essentially closed. In view of the aforementioned properties of the POCV system, we opt for the so-called {\em solution stabilization technique\/} provided by ANSYS Fluent, which is similar to the artificial-compressibility method \cite{Chimakurthi2018}.

\section{Numerical results}
\label{sec:NumRes}
In this section, we analyze the POCV on the basis of numerical investigations. Experimental tests have revealed that the POCV is dysfunctional in proportional-operation mode, in that it exhibits severe vibrations as soon as the flow rate reaches a specific setpoint, depending on the differential pressure. Experimental investigation of the origins of these vibrations is impracticable, due to the miniaturized valve size, and the perturbing effects of measurement devices. Numerical simulations therefore provide the only viable means of investigating the behavior of the POCV in proportional-operation mode, to determine the source of the vibrations.

Before applying the numerical model to investigate the behavior of the POCV in proportional-operation mode, we first regard it in the simpler settings of a fully closed configuration, and in a binary (ON/OFF) scenario. The equilibrium solution of the closed configuration also serves as initial condition for the binary and proportional-control scenarios.

\subsection{Normally closed valve}
\label{sec:NCV}
We consider the POCV in the closed configuration, for inlet pressures $p_{\text{in}}\in\{20,50,100,200,300\}$ (kPa); see Equation~\EQ{pin} and Table~\ref{tab:time-step}. The closed configuration at the various inlet-pressure levels is determined by a sequence of events. The initial configuration of the POCV corresponds to an undeformed main membrane, while the pressure difference across the channel is set to zero. In addition, the dynamic condition~\EQ{load} is disabled, so that the main membrane only deforms due to the compressed spring and the contact forces. While the main membrane deforms, the fluid domain deforms accordingly, i.e. the main membrane and the fluid are one-way coupled. After approximately $5\text{\,ms}$, the main membrane has closed and reached equilibrium. Next, the inlet pressure is set to its prescribed value, and the dynamic condition is enabled.
A two-way FSI follows, which deforms the main membrane into its equilibrium closed configuration corresponding to the imposed inlet-pressure level. The end time is set to $T=100\,\text{ms}$, at which time the main membrane has essentially reached equilibrium. 

The time-step size that is used in the simulation depends on the dynamics of the structure and the fluid which, in turn, depend on the imposed inlet pressure. The time-step size is based on the Courant number of the fluid, the mesh velocity, and the fluid and contact force magnitudes. Table~\ref{tab:time-step} summarizes the time-step size for each inlet pressure.
\begin{table}
\caption{Inlet pressures and corresponding time-step sizes for the FSI simulations.}
\label{tab:time-step}
\centering
\begin{tabular}{cccccc}
\toprule
\textbf{Pressure inlet} (kPa)& $20$&$50$&$100$ &$200$ & $300$  
\\
\midrule
\textbf{Time-step size} ($\mu$s)&$500$& 100&100 &50&10    \\
\bottomrule
\end{tabular}
\end{table}

\begin{figure}
\begin{center}
\includegraphics[scale=0.5]{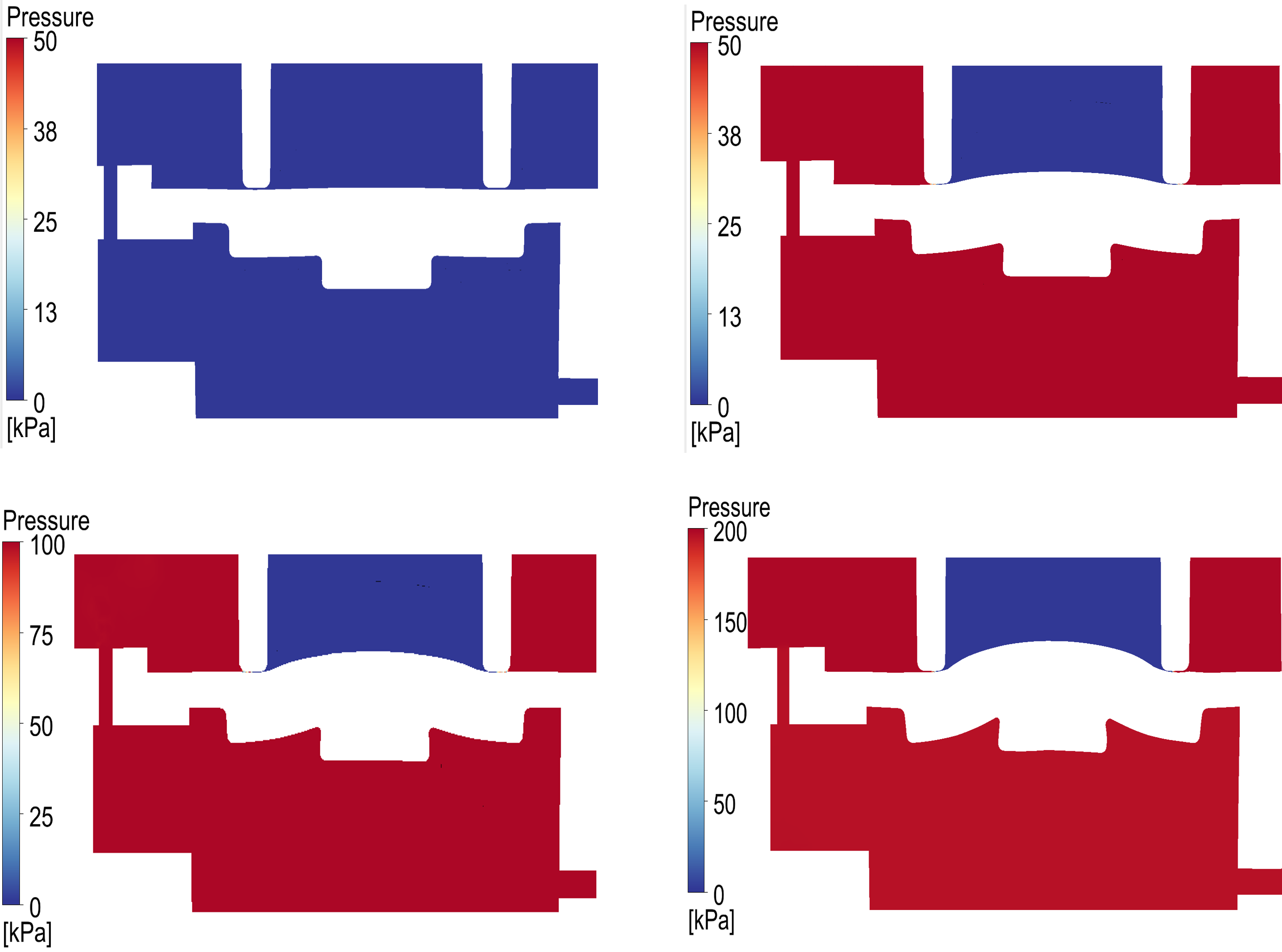}         
\caption{Plane section view of the normally closed POCV at 0 bar ({\em top left\/}), 0.5 bar ({\em top right\/}), 1 bar ({\em bottom left\/}) and 2 bar ({\em bottom right\/}). Colors indicate pressure distribution.}
\label{fig:closed-valve}
\end{center}
\end{figure}

Figure~\ref{fig:closed-valve} presents the fluid pressure 
in the closed configuration of the POCV for $p_{\text{in}}$ equal to $0,50,100$ and $200\,$kPa, in equilibrium. The fluid and solid domains are displayed in the actual deformed configuration. The plots show that the pressure is essentially piecewise constant in two disjoint regions: upstream of the main membrane, the pressure is equal to the inlet pressure, whereas downstream the pressure is equal to the outlet pressure. This implies that the gap model adequately sustains the pressure jump across the contact region. From Figure~\ref{fig:closed-valve}, one can moreover observe that the deformation of the main membrane increases as the pressure difference between the inlet and outlet increases. 

\begin{figure}
\centering
\includegraphics[scale=0.25]{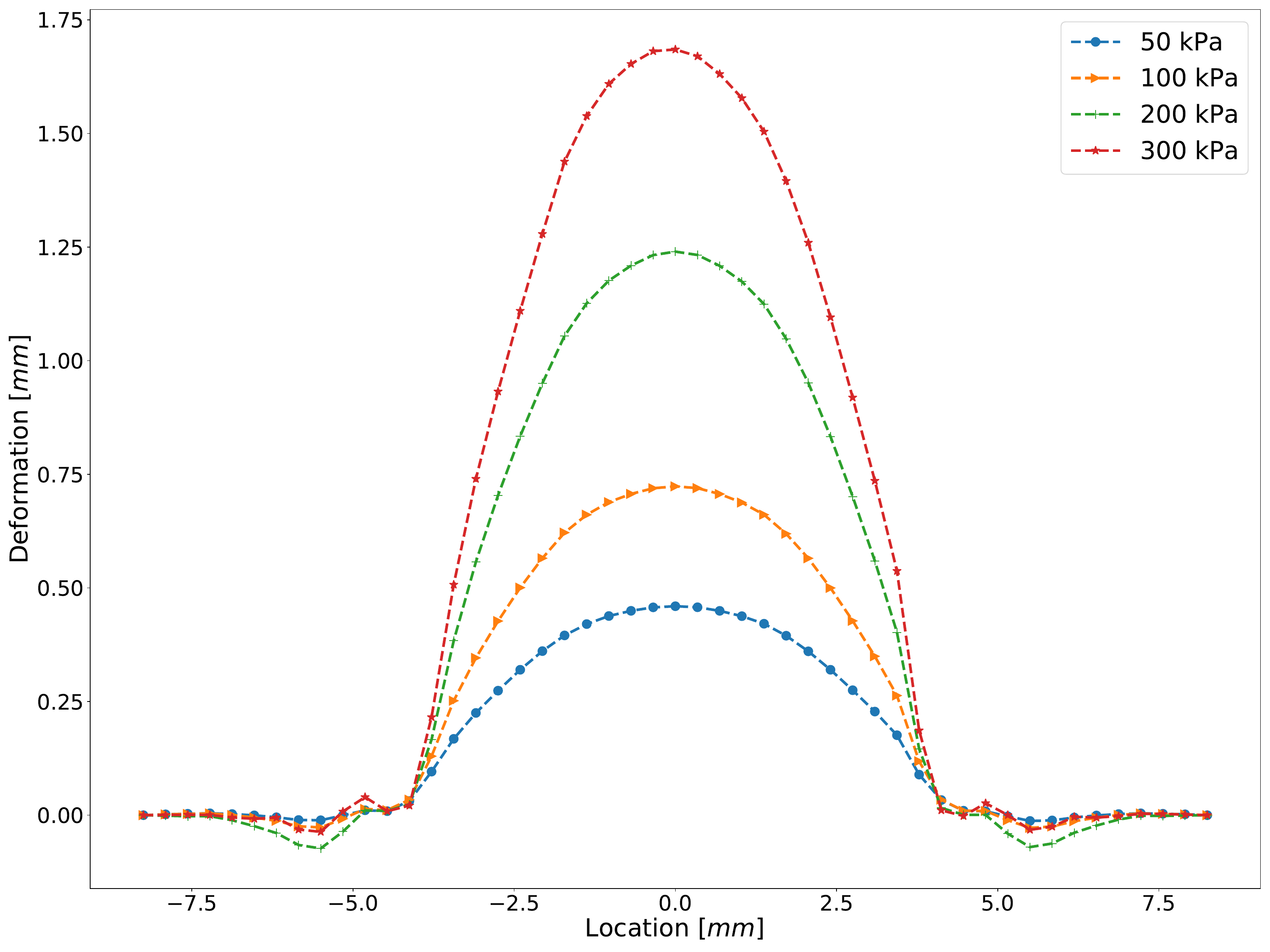}
\caption{Membrane deflection of the normally closed valve for inlet pressure values $p_{\text{in}}\in\{0.5,1,2,3\}$~bar.}
\label{fig:closed}
\end{figure}

Figure~\ref{fig:closed} plots the deformation of the top surface of the main membrane, in a cross-sectional view. One can observe that the largest displacements occur at the center of the membrane and that the deformation increases with the inlet pressure and, accordingly, the pressure difference across the main membrane. Concomitant with these high-pressure differences are large contact forces to prevent interpenetration of the membrane and valve seat or, more specifically, the dummy body. Figure~\ref{fig:closed} conveys that at an inlet pressure of $300$~kPa the implemented contact procedure is not able to completely prevent the penetration of the membrane into the dummy body in the vicinity of the valve seat, which is located at~$\pm{}5\pm[0,0.5]$\,mm. In this region, the maximum vertical displacement at $300$~kPa is approximately $0.05$\,mm. However, by virtue of the dummy body, which has thickness $0.5$\,mm, a thin but finite gap remains between the main membrane and the valve seat, which is adequate to avoid degeneration of the fluid mesh in the gap.

\subsection{ON/OFF operation}
\label{sec:digital}
The POCV is operated in a binary ON/OFF mode by means of an SMA-based actuator, which drives the pilot valve into either a closed or an open position without intermediate setpoints. The POCV working principle is then similar to that of a solenoid actuated control valve~\cite{Rajagiri_valve}. In this section, we consider numerical simulations of the opening process of the POCV. The simulation results are compared to the corresponding experimental results to assess the accuracy of the FSI model.
In practice, the opening process of the POCV proceeds according to the following stages:
\begin{enumerate}
\addtocounter{enumi}{-1}
    \item The POCV is in normally closed mode;
    \item An input signal triggers the SMA actuator;
    \item The actuator shifts the pilot membrane into the open setpoint; 
    \item The pressure in the pilot channel decreases due to the increasing flow speed;
    \item The main membrane deforms in the valve opening direction due to the changes in the pilot pressure;
    \item The main membrane position and the pressure difference determine the flow rate through the valve.
\end{enumerate}
To parallel the aforementioned stages in the numerical simulation of the valve-opening process, we proceed as follows: We start the simulation from a normally closed valve state as described in Section~\ref{sec:NCV}. Next, the pilot membrane and the adjacent fluid mesh are deformed using one-way FSI; see also Section~\ref{sec:struct}. The resulting geometry represents the open configuration of the pilot membrane. However, at this stage, the pilot bypass is still closed by means of an auxiliary wall boundary condition. In the next step, this auxiliary wall boundary in the pilot bypass is suppressed. The fluid can now flow through the pilot chamber. As a result of this flow, the pressure at the upstream wetted boundary of the main valve decreases. By virtue of the dynamic condition~\EQ{load}, the corresponding load on the main membrane deforms the membrane in the valve-opening direction. The fluid domain deforms accordingly, by virtue of the kinematic condition~\EQ{kinematic}. The contact between the valve seat and the main valve is released, and the flow can now pass directly from the upstream region to the downstream region without traversing the pilot chamber. The fluid flow and the main membrane interact in a dynamic fluid-structure-interaction process until a steady state is reached.

\begin{figure}
     \centering 
     \begin{subfigure}[b]{0.45\textwidth}
         \centering
         \includegraphics[width=\textwidth]{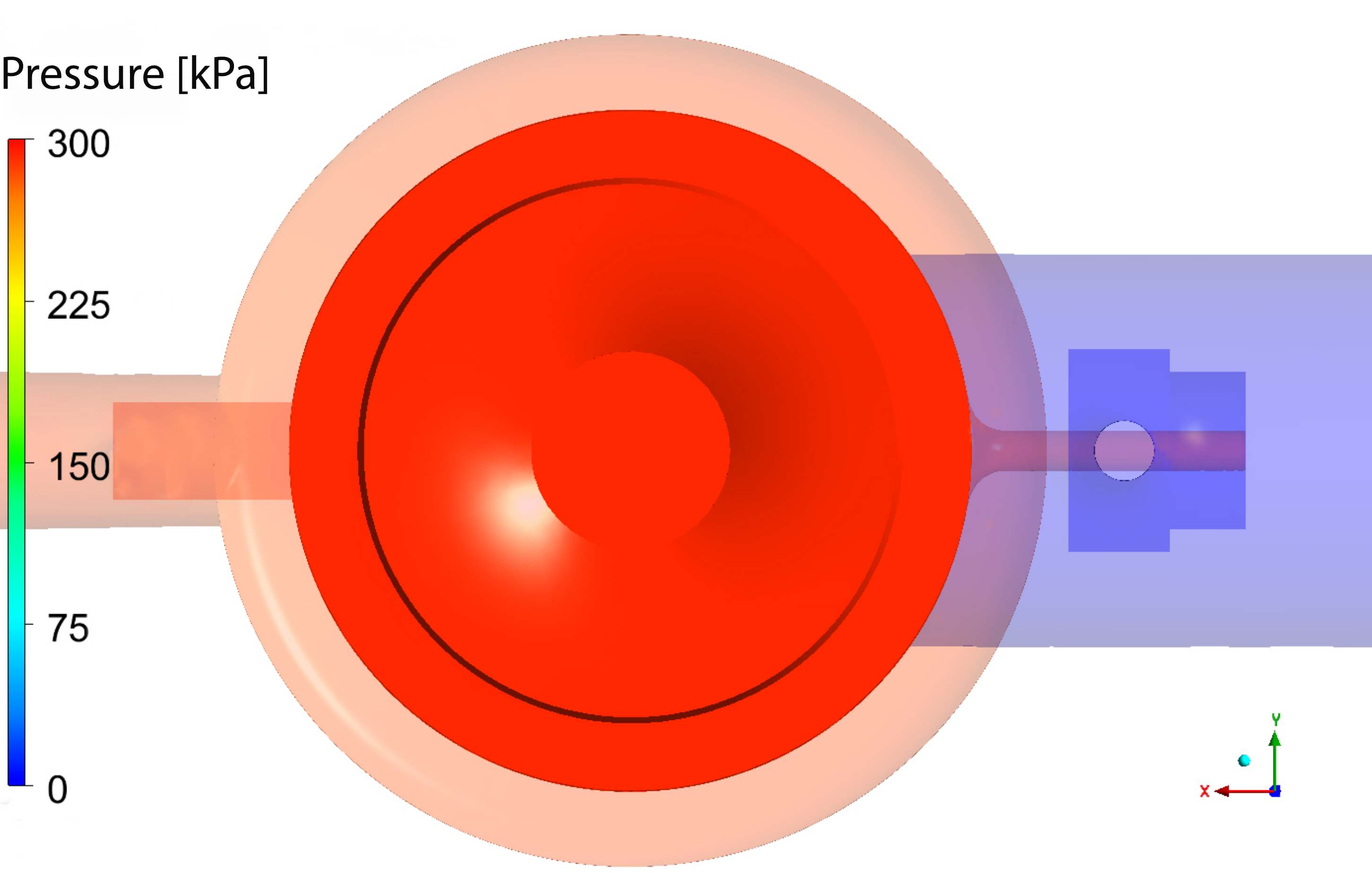}
         \caption{Fully closed valve}
         \label{fig:1}
     \end{subfigure}
     \hfill
     \begin{subfigure}[b]{0.45\textwidth}
         \centering
         \includegraphics[width=\textwidth]{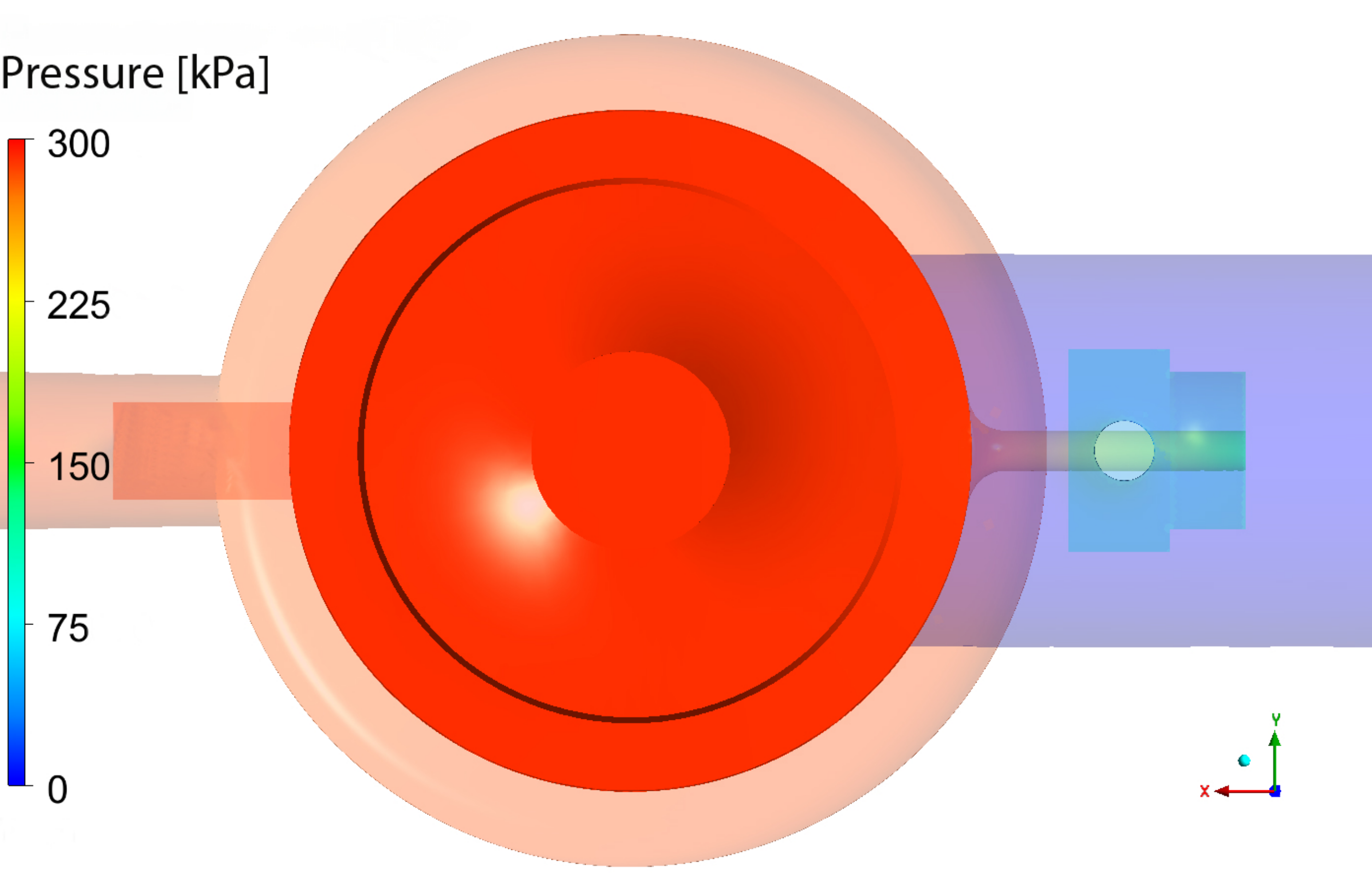}
         \caption{Pressure at 0.10005s}
         \label{fig:2}
     \end{subfigure}
          \hfill
          \newline          
    \begin{subfigure}[b]{0.4\textwidth}
         \centering
         \includegraphics[width=\textwidth]{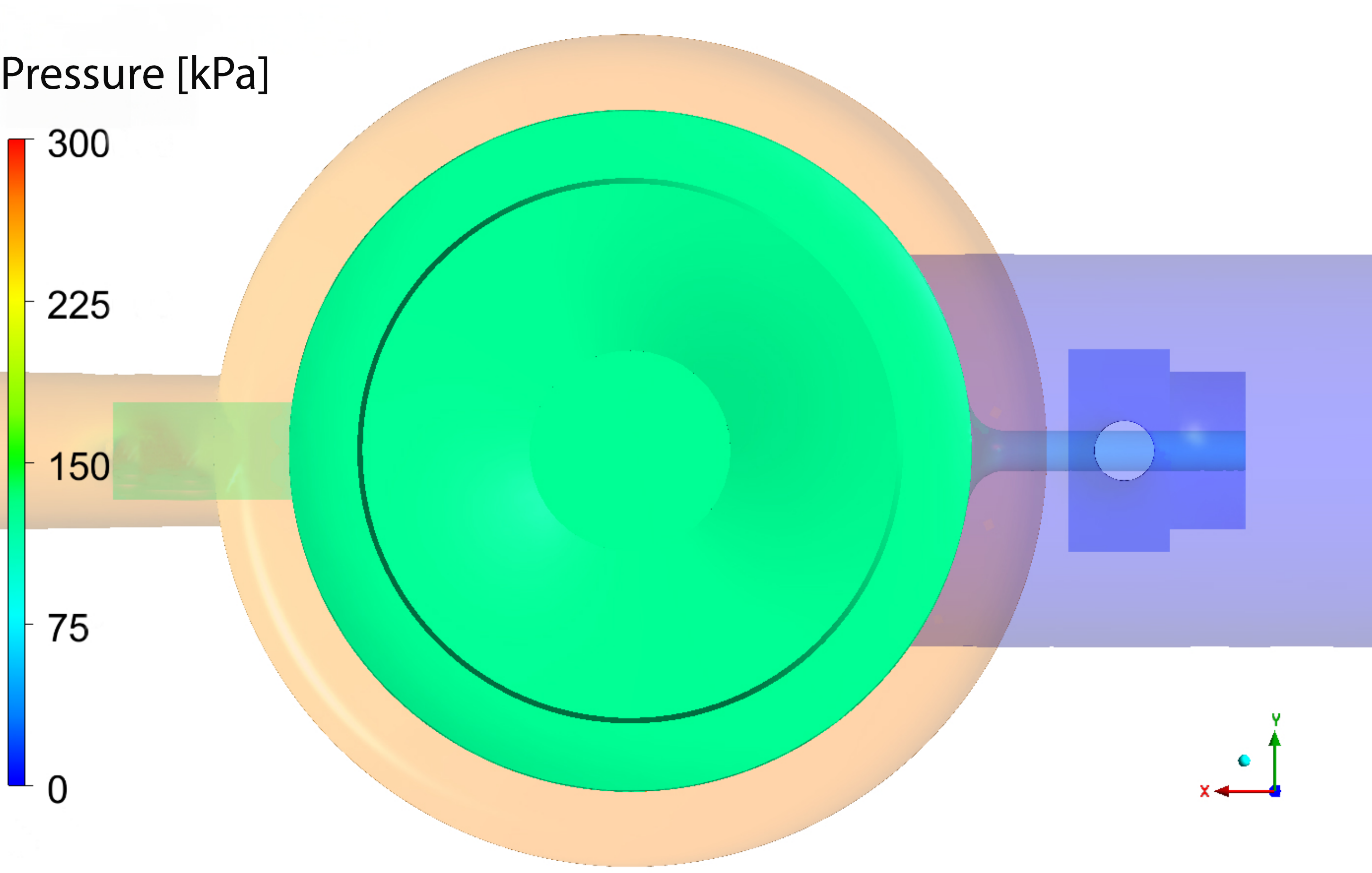}
         \caption{Pressure at 0.12s}
         \label{fig:3}
     \end{subfigure}
          \hfill  
     \begin{subfigure}[b]{0.4\textwidth}
         \centering
         \includegraphics[width=\textwidth]{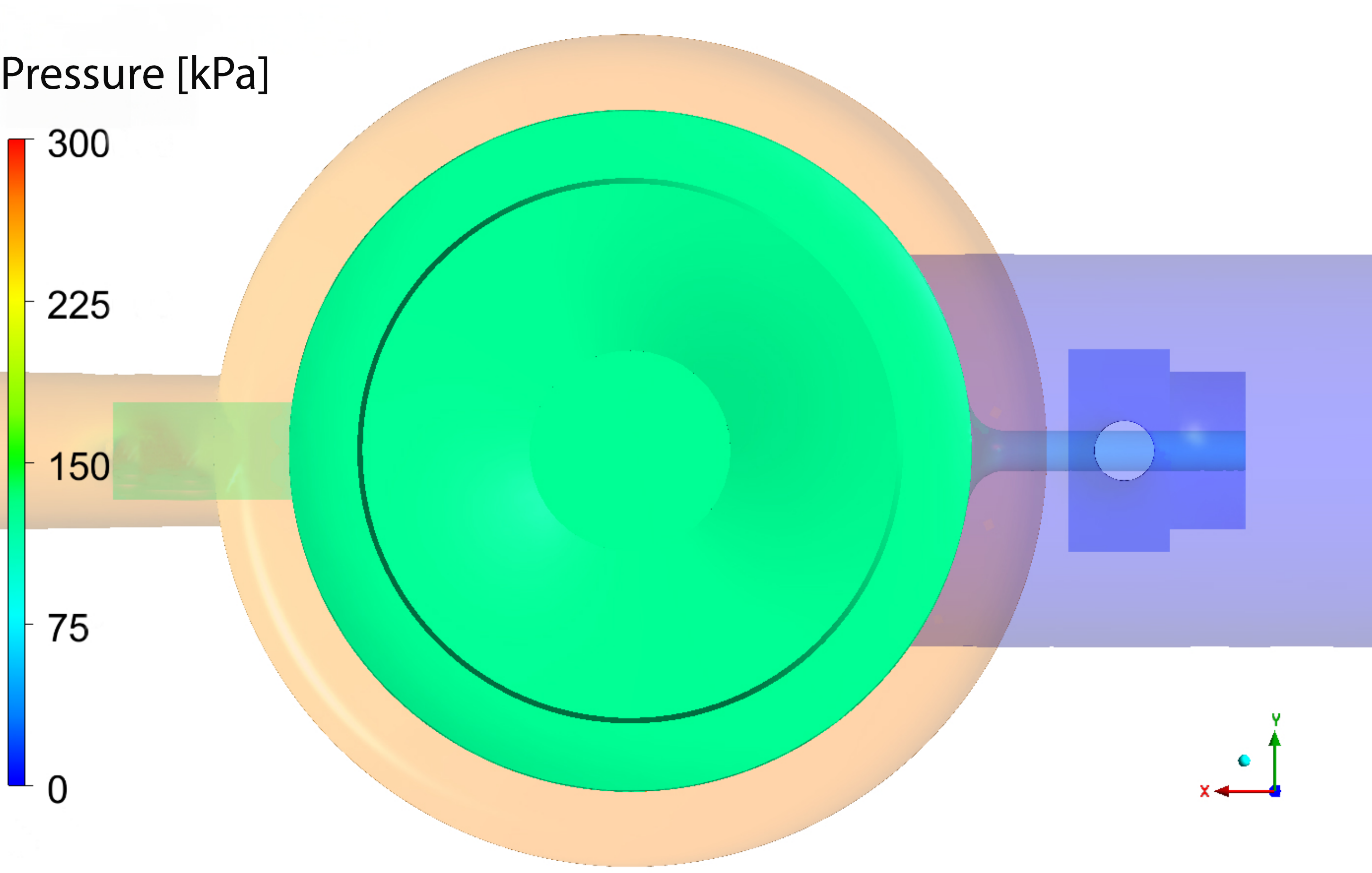}
         \caption{Pressure at 0.19995s}
         \label{fig:5}
     \end{subfigure}
  
        \caption{ Pressure contour snapshots of the POCV. The top right panel shows the pressure drop at the vicinity of the pilot membrane shortly after opening actuation, the bottom left and right panels display the pressure at different times corresponding to a fully open valve. Note the valve's fast time-to-open response from $3$ bar to ~$1.7$ bar (around $20$ ms). }
        \label{fig:pre-fsi}
\end{figure}

To demonstrate the valve's response to the opening event, Figure~\ref{fig:pre-fsi} displays four sequential snapshots of the pressure distribution across the fluid-structure interfaces and the pilot channel at various time instants. These snapshots are observed for a pressure inlet of $p_{\text{in}}=3$\,bar and display, respectively, the pressure distribution in normally-closed (NC) configuration; immediately after opening the pilot valve at~$t=10\,m$s; at $t=20\,m$s after opening the pilot valve; and after equilibration in the open position, at~$t=200\,m$s. Comparing the top-left panel and the top-right panel, one can observe that as the actuator is switched to the open position and the pilot membrane opens slightly, the pressure in the pilot tube that connects the pilot chamber to the outlet (located on the right side of the figures) immediately drops. Comparing the top-right panel to the bottom-left panel, one can observe that due to the opening of the pilot membrane, the pressure in the pilot chamber and, correspondingly, on the bottom side of the main membrane, rapidly drops from $3$\,bar to approximately $1.7$\,bar, in less than $20$ ms. The pressure drop induces a change in the load exerted on the main membrane, resulting in the main membrane undergoing deformation and transitioning into a fully open configuration (bottom-right panel). The open configuration represents a new equilibrium state of the valve system which depends on the applied inlet pressure.

\begin{figure}
\begin{center}
\includegraphics[width=\textwidth]{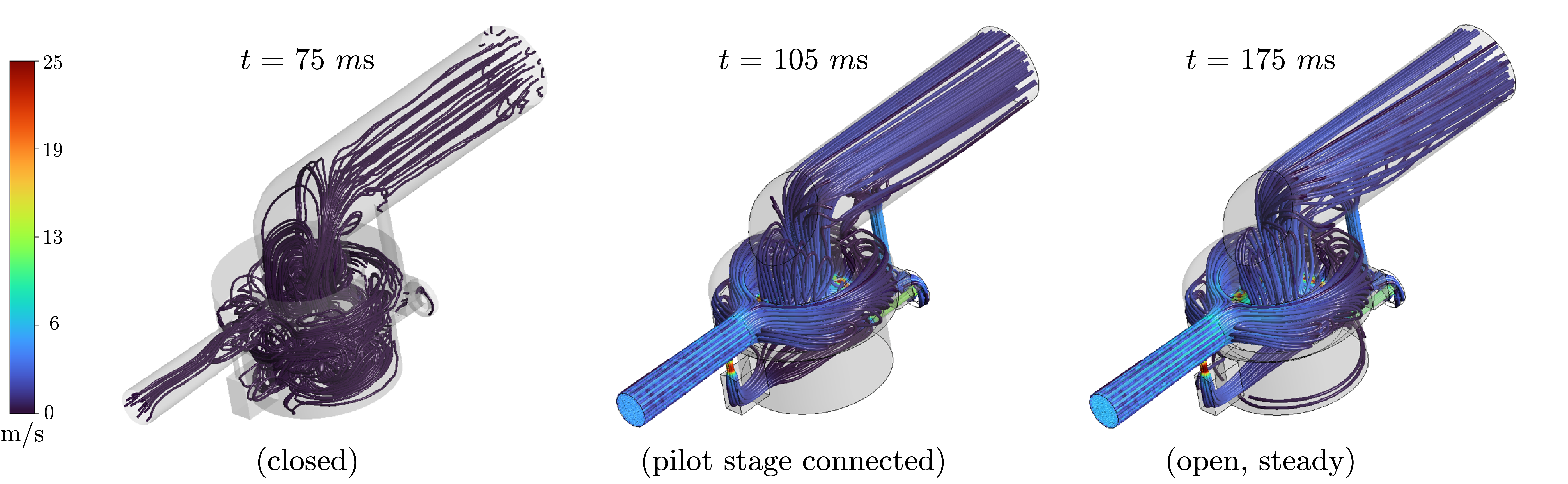}
\end{center}
\caption{Streamlines during the opening of the microfluidic POCV at $t=75$\,ms ({\em left\/}), $t=105$\,ms ({\em center\/}) and $t=175$\,ms ({\em right\/}) for an inlet pressure of $3$\,bar.}
\label{fig:STR-valve}
\end{figure}

Figure~\ref{fig:STR-valve} shows the streamlines of the flow in the POCV at different time instances, for an inlet pressure of $3$\,bar. At $t=75$\,ms, the POCV is still closed and the fluid velocity essentially vanishes throughout the entire POCV. At $t=105$\,ms, shortly after the opening of the pilot membrane, the fluid exhibits significant flow velocities, primarily near the valve seat and along the tube that connects the pilot valve to the main valve. The right-most panel of Figure~\ref{fig:STR-valve} corresponds to the steady state of the flow in a fully open configuration. Noticing the similarity of the streamline patterns in the center and right-most panels, one can infer that after opening, the flow very quickly settles into a stable, steady regime. In the open configuration, most streamlines connect the outlet to the inlet via the gap between the main membrane and the valve seat, and only a few pass through the pilot stage. This indicates that in the open configuration, the secondary flow via the pilot stage is limited. Noting that in the open configuration, the velocity at the inflow and outflow regions is essentially uniform, the simulation results predict a smooth and stable flow at the outlet. 

\begin{figure}
\begin{center}
\includegraphics[width=0.75\textwidth]{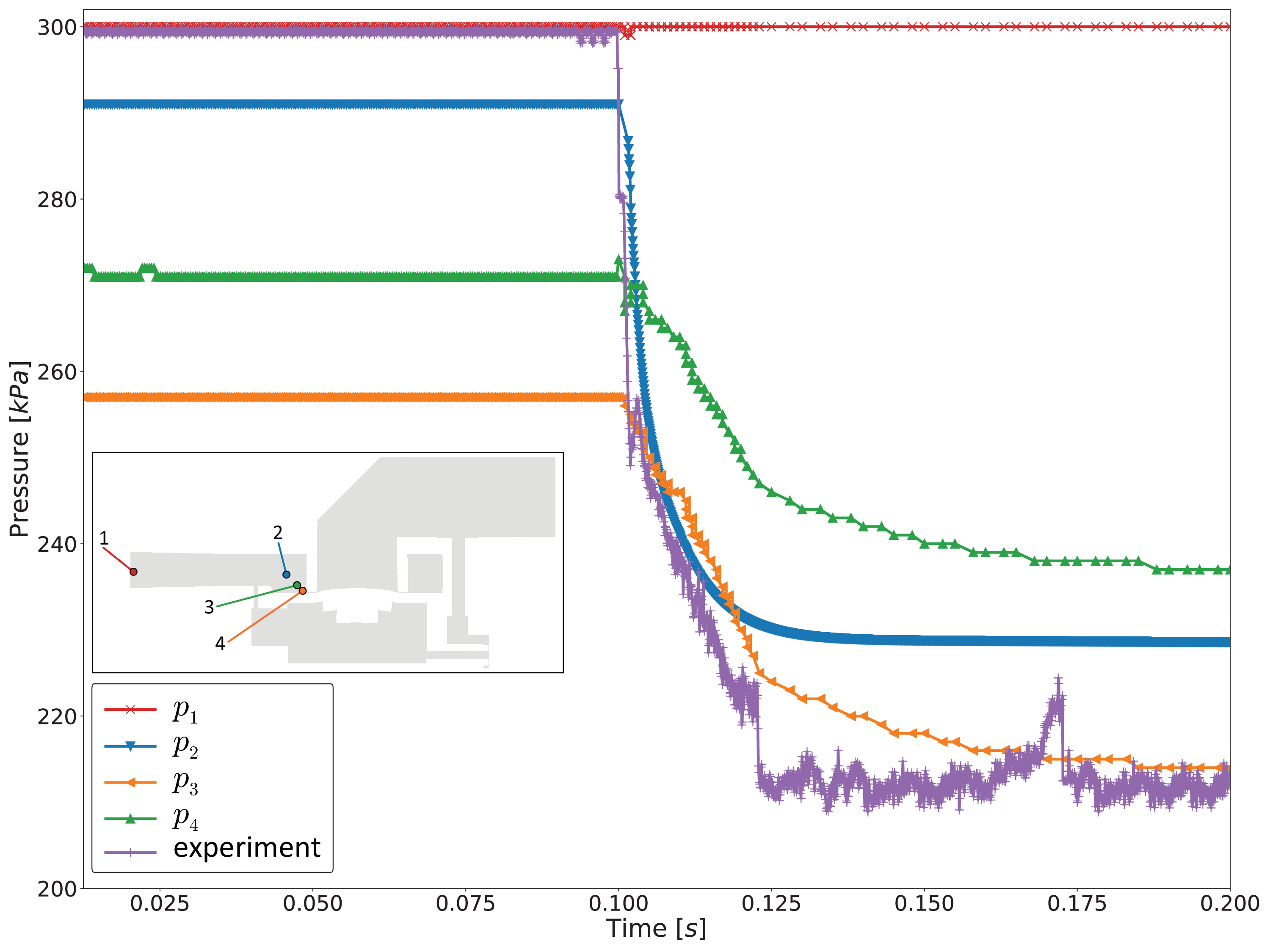}    
\end{center}
\caption{Pressure evolution during opening of the POCV at differential pressure $3$\,bar: Computed pressure at 3 points near the valve-seat gap and at the inlet, and experimental data.}
\label{fig:validation}
\end{figure}

To validate the results of the numerical simulations in relation to the response of the actual POCV device, Figure~\ref{fig:validation} presents a comparison between the computed and experimentally obtained pressure differences across the valve, for a differential pressure of $3$\,bar. In view of the uncertainty of the exact location of the pressure measurement in the experimental setup, e.g. due to the size of the probe, in the simulation the pressure has been probed at three representative distinct, nearby locations. These locations are indicated in the inset in Figure~\ref{fig:validation}. The computed pressure at the inlet indeed coincides with the imposed inlet pressure of~$3$\,bar. In the closed configuration, the pressure is not uniform in the inlet region, but instead, it decreases as the gap between the main valve and the valve seat is approached. This suggests that there is some leakage through the gap in the numerical model. Bearing in mind the small dimensions near the gap, minute flow velocities can lead to significant variations in the pressure. It is notable that the experimentally obtained pressure is also non-uniform: in the closed configuration, the pressure at the probe drops to $298$\,kPa, i.e. $2$\,kPa less than the imposed inlet pressure of $300$\,kPa. This suggests that the experimental setup also exhibits some minute leakage. The computed results indicate that the steepness of the pressure drop during the opening of the main valve and the post-opening pressure level depends sensitively on the location. In addition, the results indicate that the pre-to-post opening pressure drop is non-uniform with respect to the distance to the valve-seat gap.
The computed post-opening pressure level at the measurement point closest to the gap agrees well with the experimental data. However, the pre-opening pressure level at this location is significantly lower than in the experiments. The strong sensitivity of the pressure readings to the location and to the leakage through the valve-seat gap, evident from the simulation results, in combination with the uncertainty in the location of the measurement device, indicates that the pressure-drop comparison in Figure~\ref{fig:validation} can only serve a qualitative purpose.

\begin{figure}
\begin{center}
\includegraphics[width=0.75\textwidth]{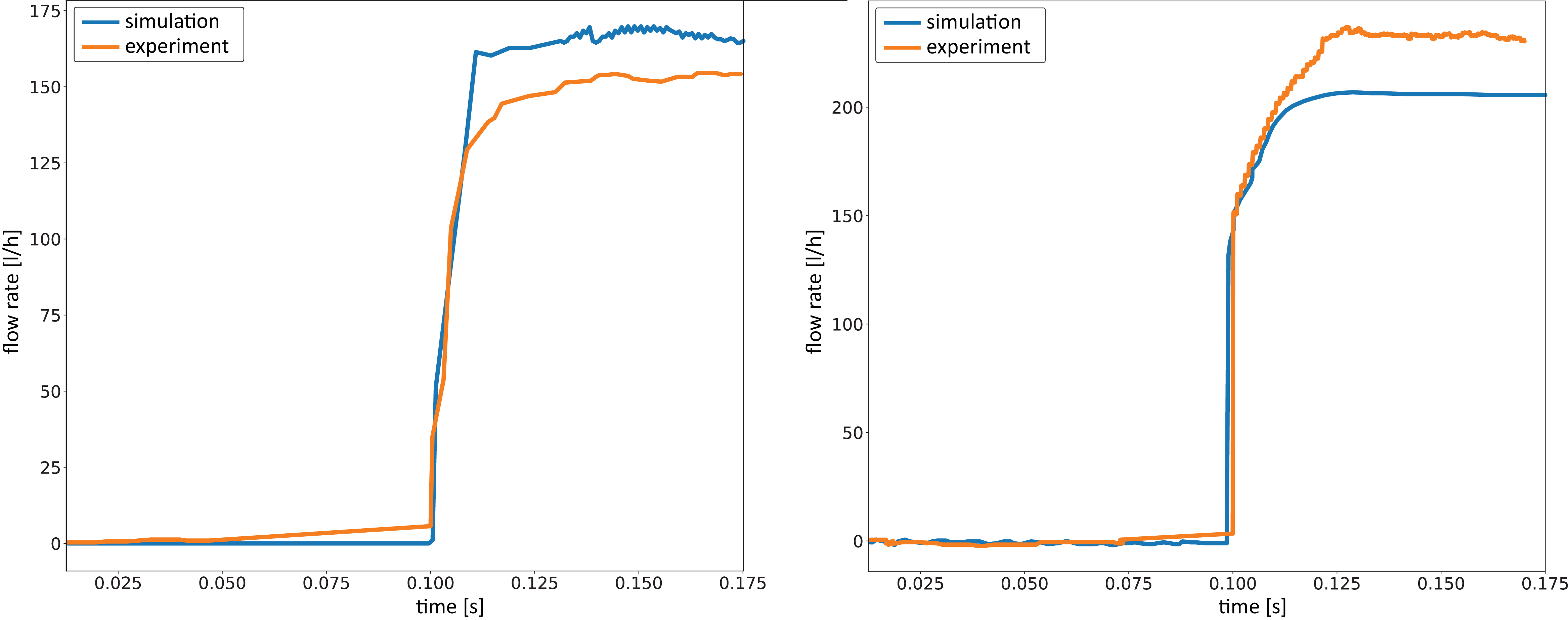}    
\end{center} 
\caption{Computed flow rate and experimentally obtained flow rate for a differential pressure of $2$\,bar ({\em left\/}) and $3$\,bar ({\em right\/}).}
\label{fig:mfr}
\end{figure}

To further assess the agreement between the numerical model and the actual POCV device, Figure~\ref{fig:mfr} presents the temporal variation of the flow rate through the POCV in ON/OFF operation for differential pressures of $2$\,bar and $3$\,bar. For the inlet pressure of $2$\,bar (resp. $3$\,bar), the sudden increase in the flow rate from an initial value of $0$\,l/h to approximately $150$\,l/h (resp. $200$\,l/h), which occurs immediately after the control pilot valve is opened, inducing the opening of the main valve, aligns well with the experimental results. 
For inlet pressure $3$\,bar, the computed result closely follows the more gradual increase in the flow rate in the time interval $100-120\,m$s, exhibited by the experimental data. For the experimental results, the increase of the flow rate however extends to a slightly longer time interval than for the computed results. Consequently, the experimentally observed equilibrium flow rate exceeds the computed flow rate by approximately 10\%. Conversely, for the $2$\,bar case, the computed equilibrium flow rate exceeds the experimental flow rate by approximately 10\%. In view of the sensitivity of the flow rate to the details of the valve deformation near the valve seat, we assess the overall agreement between the computed and experimental flow rate evolutions as good.

\subsection{Proportional-control operation}
The considered POCV in principle allows to be operated as a proportionally controlled valve, providing the option to dynamically adjust the flow as required by the user by means of an analog input signal instead of the binary signal that is used to operate the valve in ON/OFF mode. 
Noting that the POCV is controlled by a voltage applied to the SMA-based actuator, the analog signal can assume intermediate values in the voltage range $[0,V_{\mathrm{max}}]$, where $V_{\mathrm{max}}$ corresponds to the voltage at which the control valve is fully open. Proportional operation greatly extends the operational range of the microfluidic POCV to, for instance, dosing and mixing tasks that require precise flow modulation. However, the POCV under investigation is deficient in proportional-control operation, in that the device exhibits severe vibrations once the differential pressure exceeds a certain threshold. In this section, we apply the developed FSI model to examine the vibrations in the POCV in proportional-control mode.

\begin{figure}
\centering
\includegraphics[scale=0.2]{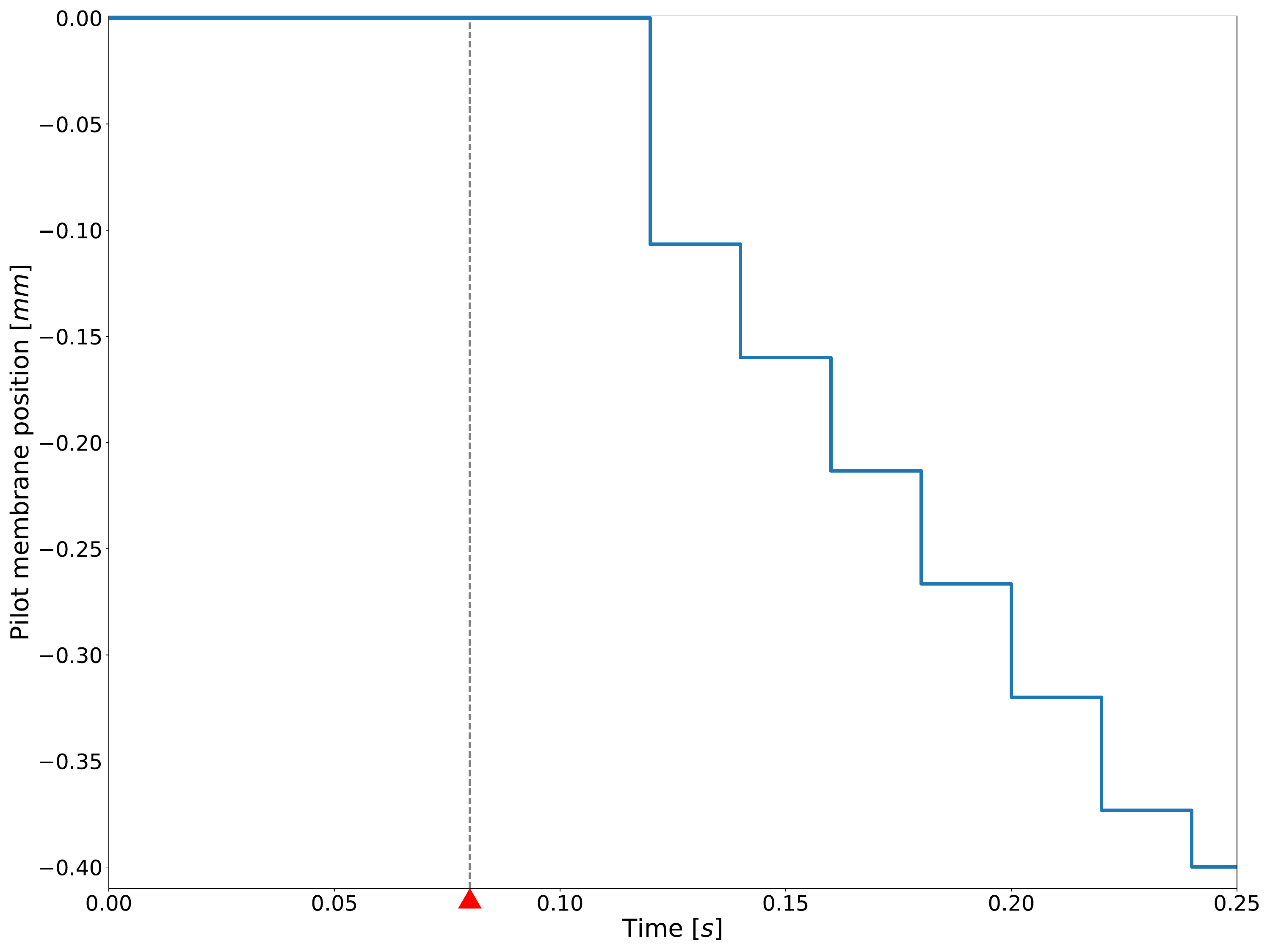}
\caption{Step-wise displacement of the pilot membrane in the proportional-control scenario of the POCV. The opening process follows three steps: ($1$): Establishing the closed state of the valve system until $t=0.08$\,s ({\em red triangle\/}); ($2$): Suppressing the wall boundary condition that models the closed pilot spool and establish setpoint = $20\,\mu$m for $t\in(0.08,0.12)$\,s; ($3$): opening of the control valve following the imposed step-wise displacement of the SMA-actuator mount point for function $t\in(0.12,0.125)$\,s.
The valve displacement of $-0.4\,m$m corresponds to the fully open setpoint in the ON/OFF scenario.}
\label{fig:stp-wise}
\end{figure}

In Section~\ref{sec:digital} it was shown that after opening, the flow in the POCV very quickly equilibrates. As the envisaged proportional-control operation is relatively slow compared to this equilibration time, the proportional-control operation can essentially be regarded as a sequence of quasi-stationary operations. To facilitate the analysis, we therefore consider a step-wise progressive opening (negative displacements) of the control valve, as depicted in Figure~\ref{fig:stp-wise}. 
In view of the minimal gap that must be retained to accommodate the fluid mesh, an initial gap of approximately $20\,\mu$m is applied, corresponding to $5$\% of the displacement of the pilot membrane in the fully open position. However, the initial gap is closed by means of an auxiliary wall boundary condition; see also Section~\ref{sec:digital}. At $t=0.08$\,s, this auxiliary condition is suppressed, and the pilot valve essentially opens to $20\,\mu$m, without displacement of the pilot valve. The opening is subsequently step-wise increased in accordance with Figure~\ref{fig:stp-wise}, by imposing a displacement on the pilot valve at the mounting area of the SMA-based actuator. Let us note that the maximum negative displacement at $-0.4\,m$m corresponds to the opening of the pilot valve considered in the ON/OFF scenario in Section~\ref{sec:digital}.

\begin{figure}
\centering
\includegraphics[width=0.75\textwidth]{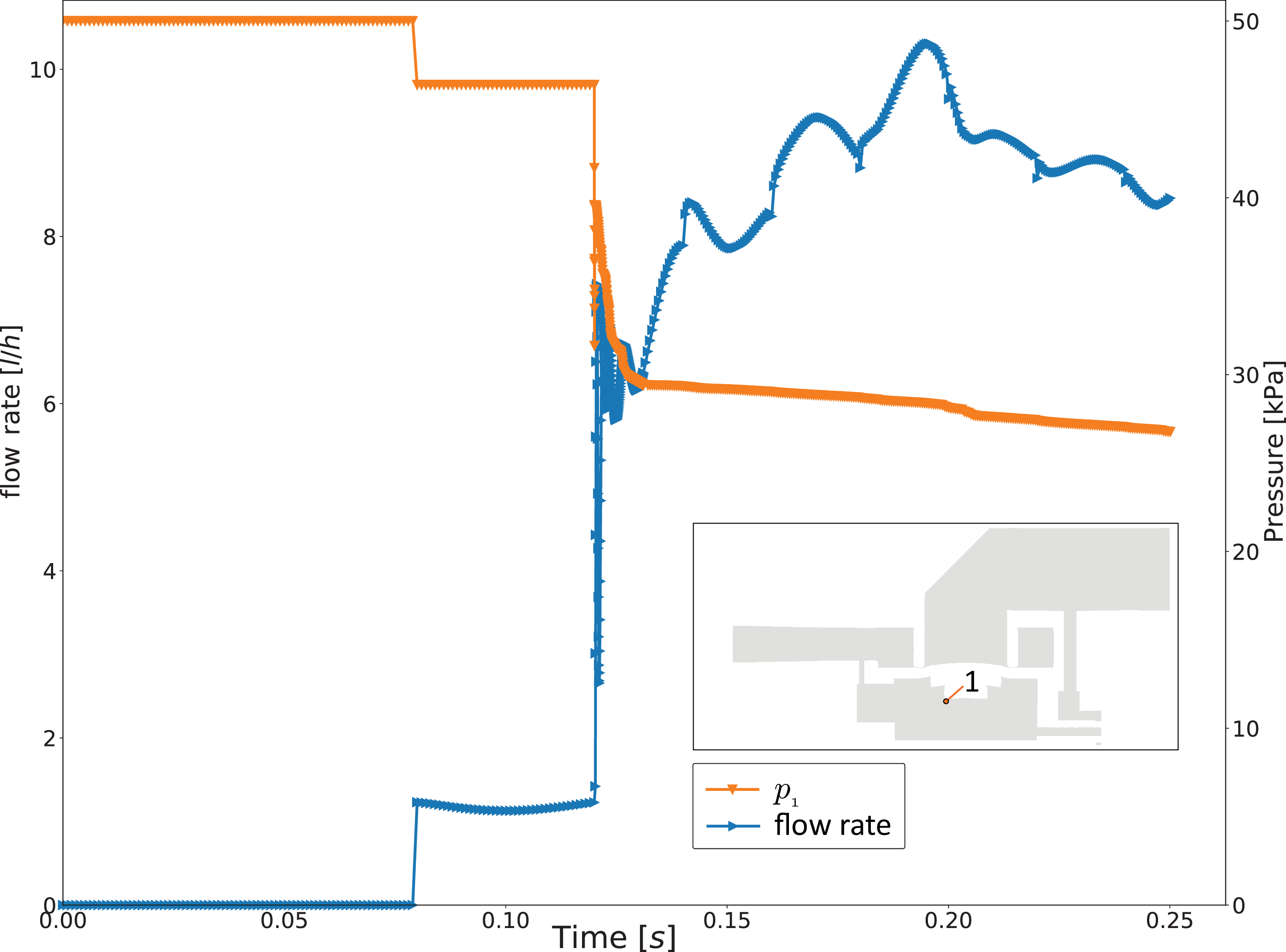}    
\caption{Flow-rate and pressure evolution during the opening of the POCV in accordance with Figure~\ref{fig:stp-wise} at differential pressure $0.5$bar.}
\label{mfr-pr}
\end{figure}

\begin{figure}
\begin{center}
\includegraphics[width=\textwidth]{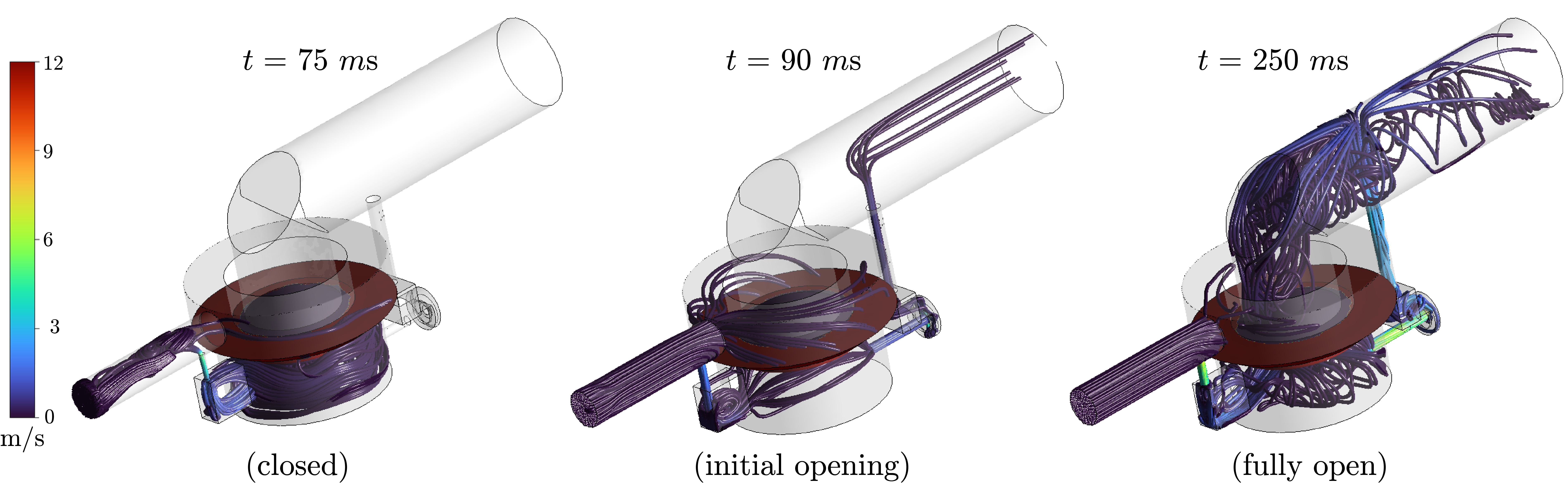}
\end{center}
\caption{Streamlines during opening of the POCV according to Figure~\ref{fig:stp-wise} at $t=75$\,ms ({\em left\/}), $t=90$\,ms ({\em center\/}) and $t=250$\,ms ({\em right\/}) at differential pressure $0.5$\,bar.}
\label{fig:Prop-opening0.5}
\end{figure}

To establish that a minimal pressure differential is required for the POCV to function, we first regard an inlet pressure of $0.5$\,bar. Figure~\ref{mfr-pr} displays the evolution of the flow rate across the outlet and the pressure at a specific point within the pilot chamber during the proportional opening scenario in accordance with the step-wise displacement of the pilot membrane; see Figure~\ref{fig:stp-wise}. One can observe that the opening of the pilot membrane at $t=0.08$\,s, via removal of the auxiliary wall boundary condition, induces a pressure drop of approximately $5\,k$Pa. In addition, the flow rate instantaneously increases to $1.5\,$l/h. The first step in the displacement of the pilot membrane at $0.12$\,s again yields a sharp decrease in the pressure in the pilot chamber, to approximately $30\,k$Pa. Simultaneously, the flow rate increases gradually to approximately $8\,$l/h.  The subsequent step-wise increments of the displacement of the pilot membrane, bear only a moderate effect on the pressure in the pilot chamber, and a nonuniform effect on the flow rate. At the maximum displacement of the pilot membrane at $t=0.25$\,s, the pressure in the pilot chamber (at the measurement point) is approximately $27\,k$Pa, while the flow rate levels off at approximately $8.5$\,l/h. It is notable that this flow rate is disproportionally lower than the observed flow rates for $2$ and $3$\,bar in Section~\ref{sec:digital}. To further elucidate the operation of the POCV at $0.5\,$bar, we regard in Figure~\ref{fig:Prop-opening0.5} snapshots of the velocity streamlines before opening, at $t=0.075$\,s, immediately after opening, at $t=0.09$\,s, and at full opening, at $t=0.25$\,s. One can observe that after opening the pilot valve, the streamlines pass through the pilot bypass only. This indicates that the pressure differential of $0.5$\,bar is insufficient to unseal the main valve. This also explains why the flow rate at $0.5$\,bar is disproportionally less than the flow rate for $2\,$bar and $3\,$bar observed in Section~\ref{sec:digital}, and the unsteadiness of the flow in the intermediate and fully open configuration.

\begin{figure}
\begin{center}
\includegraphics[width=0.8\textwidth]{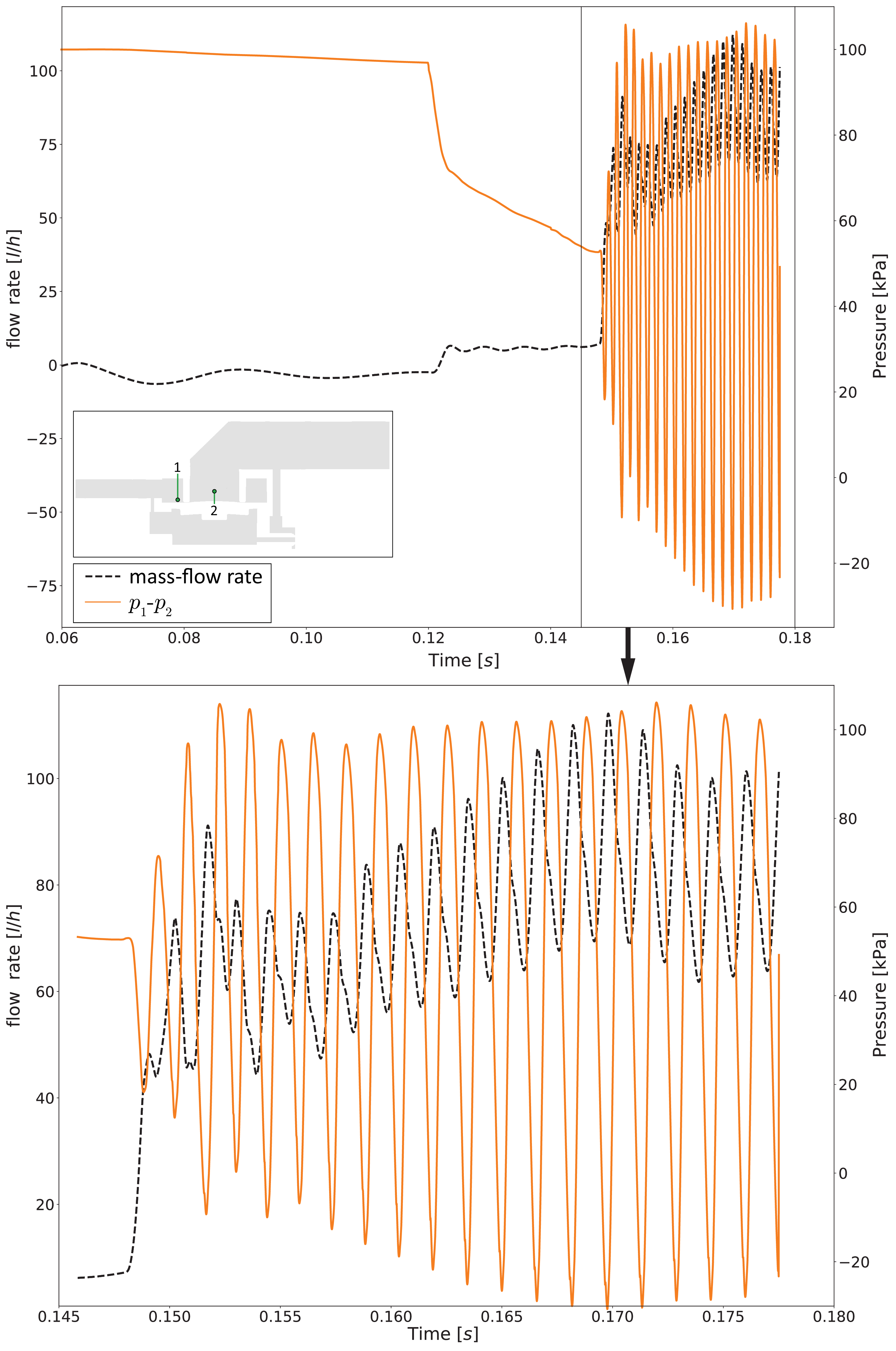}
\end{center}
\caption{Evolution of the flow rate and upstream-downstream pressure difference during proportional operation of the POCV for an inlet pressure of~$1\,$bar.}
\label{fig:oscillations}
\end{figure}

In the remainder of this section, we focus on the proportional operation of the POCV for a pressure differential that suffices to unseal the main valve, to examine the vibrations in the POCV in proportional-control mode. To illustrate the valve response after the main valve opening, 
Figure~\ref{fig:oscillations} plots the evolution of the difference in the pressure upstream and downstream of the main valve, and the flow rate across the outlet, during the proportional valve opening in accordance with Figure~\ref{fig:stp-wise}, for an inlet pressure of $1$\,bar. Until the opening of the pilot valve at $t=0.08$\,s, the upstream pressure is essentially uniform and equal to the specified inlet pressure. Accordingly, the flow rate vanishes, except for some minute leakage through the valve-seat gap. The pilot valve is opened to $5\%$ at $t=0.08$\,s by suppressing the auxiliary wall boundary condition near the pilot membrane. The pressure difference across the main membrane caused by the $5\%$ opening of the pilot membrane does not suffice to open the main valve. Hence, the upstream-downstream pressure difference does not change, and the flow across the POCV remains negligible. The pilot valve displacement at $t=0.12$\,s to $25\%$ open yields an abrupt reduction in the upstream-downstream pressure difference, followed by a linear decay, and induces a flow of approximately $5$\,l/h across the POCV. At this setting of the pilot valve, the main valve is still closed, and the flow passes through the pilot conduit; see also Figure~\ref{fig:streamlines_osc} below. Further displacement of the pilot valve at $t=0.14$\,s to an approximately $38\%$ open setpoint leads to a sharp increase in the flow rate from $5$\,l/h to $45$\,l/h, and a concurrent sharp reduction of the upstream-downstream pressure difference from $53\,k$Pa to $19\,k$Pa. These sharp changes in the flow rate and pressure difference portend severe oscillations in the flow rate and pressure difference. One can observe that the oscillations exhibit a consistent frequency, with some modulations. The oscillations occur concurrent with the opening of the main valve, which indicates that the considered POCV cannot be used in proportional control mode. For completeness, we mention that the FSI simulation terminated after $t=0.17752$\,s due to failure to converge in the subsequent time step.

To investigate the oscillatory behavior of the POCV in proportional-control mode, we regard the flow in the POCV during proportional opening. Figure~\ref{fig:streamlines_osc} displays the streamlines at different time instances. The top-left panel displays the streamlines in the POCV at $t=128.8\,m$s, shortly after opening the pilot valve to $25\%$. One can note that the streamlines between the inflow and outflow pass through the pilot channel only, not through the valve-seat gap, indicating that the main valve is still closed. The remaining panels in Figure~\ref{fig:streamlines_osc} display snapshots of the streamlines during the oscillation of the POCV. The top-right panel displays the streamlines at $t=160.4\,m$s. This moment corresponds to a trough in the differential pressure; see Figure~\ref{fig:oscillations}. The troughs in the differential pressure coincide with moments at which contact between the main valve and the valve seat is established, i.e. when the main valve closes. Simultaneously, a peak in the flow rate occurs. One can observe that at $t=160.4\,m$s, no streamlines pass from the inlet to the outlet via the valve-seat gap, in accordance with the fact that the main valve is closed.
The bottom-left panel displays the streamlines at $t=160.8\,m$s. At this instant, the valve is still closed, the differential pressure is increasing, and the flow rate is decreasing. The streamlines exhibit a similar pattern as for $t=160.4\,m$s, in accordance with the fact that the main valve is still closed. The bottom-right panel displays the streamlines at $t=161.4\,m$s. In this plot, multiple streamlines pass from the inlet to the outlet through the valve-seat gap, indicating that the main valve is open. From the color of the streamlines, one can observe that the flow velocities in the vicinity of the gap are very large.
With reference to Figure~\ref{fig:oscillations}, we note that the flow rate exhibits a local minimum at $t=161.4\,m$s.
\begin{figure}
\begin{center}
\includegraphics[width=\textwidth]{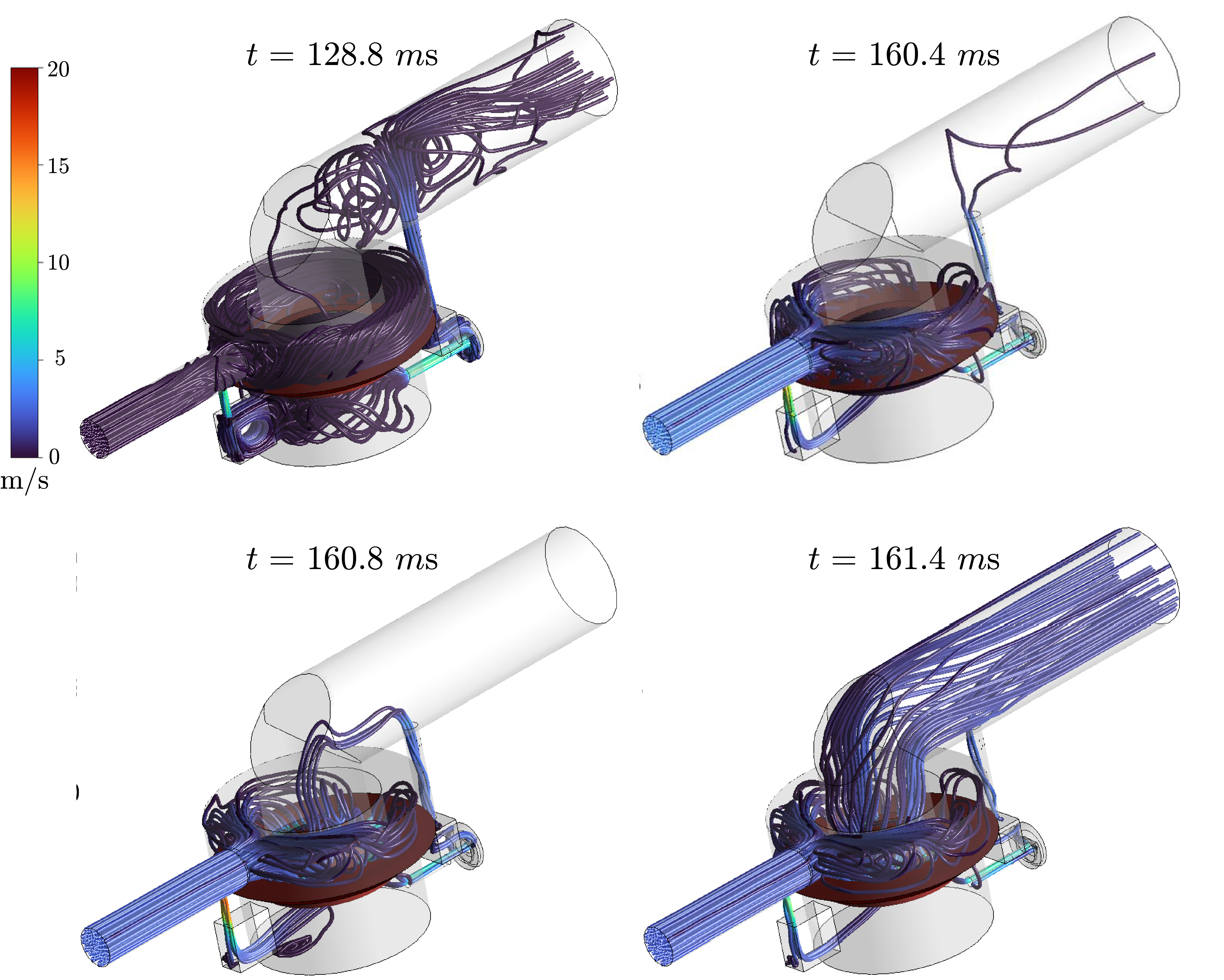}
\end{center}
\caption{Streamlines during attempted proportional operation of the POCV at a differential pressure of $1$\,bar 
at $t=128.8\,m$s ({\em top-left\/}), $t=160.4\,m$s ({\em top-right\/}), $t=160.8\,m$s ({\em bottom-left\/}) and $t=161.4\,m$s ({\em bottom-right\/}).}
\label{fig:streamlines_osc}
\end{figure}

Closer inspection of the results in Figures~\ref{fig:oscillations} reveals that during the oscillation, troughs in the differential pressure coincide with peaks in the flow rate. These extrema correspond to moments at which the main valve closes. Conversely, pressure peaks occur at moments at which the main valve opens (results not displayed). It is further noteworthy that pressure peaks do not coincide with local minima in the flow rate, which indicates that the oscillation is not simply harmonic. Although it appears contradictory that a significant outflow can be generated while the main valve is closed, one has to bear in mind that the main membrane exhibits significant deformations during the oscillation cycle, including the part of the cycle in which the valve is closed. Hence, during this part of the cycle, the volume that exits the POCV is balanced by the motion of the main membrane. Similarly, it is noteworthy that the flow rate continues to decrease when the main valve opens. This can be interpreted as that part of the volume that passes through the gap, is stored in the volume which emerges from the motion of the main membrane.

A main mechanism underlying the dynamic instability of the POCV in (attempted) proportional operation, appears to be the significant pressure drop that occurs in the thin valve-seat gap after opening. At the transition from the thin valve-seat gap to the outlet channel, in the vicinity of the main membrane, the flow sustains a strong divergence, which results in a significant pressure drop, as well as strong vorticity; see Figure~\ref{fig:streamlines_osc}. The strong pressure drop in the valve-seat gap counteracts the opening of the membrane. Moreover, if the motion of the membrane is (locally) reversed due to the negative pressure in the valve-seat gap, the Venturi effect~\cite{Opstal:2013fk,timo_thesis} becomes operative, causing the valve-seat gap to collapse. The Venturi effect pertains to a negative pressure singularity  ($p\to-\infty$) that occurs in a fluid between two approaching surfaces; see~\cite[Thm.29]{timo_thesis}. Hence, the collapse of the valve-seat gap occurs forcefully, which is consistent with the strong oscillations that have been observed in experimental investigations of the POCV in proportional operation.
The above elaboration of the mechanism that is responsible for the oscillatory response of the POCV in proportional operation, suggests that an important condition for successful, non-oscillatory operation of the POCV is that during the opening of the main valve, the thin-gap regime is transcended sufficiently fast. In the ON/OFF-operation test cases considered in Section~\ref{sec:digital}, the instantaneous complete opening of the control valve yields a pressure drop across the main membrane that apparently is sufficiently strong to fulfill this condition.






\section{Conclusion}
\label{sec:Concl}
Valves play a central role in hydraulic and microfluidic devices. 
Extending functionality of valves beyond their original operational range, can provide an important contribution in miniaturization and 
cost reduction. From that perspective, in the present paper we have numerically investigated a smart pilot-operated control valve (POCV) that has originally been designed for ON/OFF switching, with the aim of elucidating vibrations that have been observed experimentally in proportional operation of the device. The numerical investigations consider the fluid-structure-contact interaction that occurs in the POCV during operation.

For the fluid-structure-interaction simulations, we have developed a numerical model in the ANSYS framework, with the auxiliary aim of testing to what extend such a generic simulation environment is suitable for simulating a complex industrial FSI problem. On account of the incompressibility of the considered fluid and the relatively high density of the fluid, the considered POCV problem represents a strongly coupled FSI problem which, in addition, displays characteristics of the incompressibility dilemma, because in the closed configuration, the upstream volume is essentially closed. The so-called {\em solution-stabilization technique\/} provided by ANSYS-Fluent, which is similar to artificial compressiblity, allows to stabilize the iterations in the partitioned solution strategy.

The numerical results convey that the valve exhibits reliable sealing properties and allows fast switching from open to closed positions. The simulations show that the considered POCV is reliable in ON/OFF control operations. The corresponding numerical results display good agreement to corresponding experimental data, in terms of flow rate and pressure difference. Investigation of the proportional-operation regime, conveys that the POCV is unsuitable for proportional control, and is susceptible to vibrations in this regime. Self-excited vibrations occur as soon as the main valve is unsealed, resulting in membrane oscillations and pressure pulsations. The numerical results suggest that the vibrations are essentially caused by the sharp pressure drop that occurs in the valve-seat gap, which counteracts the opening of the main valve, and the ensuing Venturi effect, which causes a forceful collapse of the valve-seat gap. An important condition for successful operation of the POCV appears to be that during opening of the main valve, the thin-gap regime is transcended sufficiently fast.

\newpage
 \appendix
\section{Membrane material characterization}
\label{calib}
 \subsection{Constitutive hyperelastic material model}
 \label{constitutive_hyp}
Experimental data are required for the appropriate calibration of a hyperelastic constitutive model. In this context, three uni-axial tensile tests were performed in the DICAr department at the University of Pavia \cite{Compmech}. The samples' dimensions are shown in the table \ref{tab:dimension}. The MTS Insight 10 kN testing device was used to load the specimen with a $250$ N tensile load.\\  
\begin{table}[H]
\caption{The specimen dimension in terms of length, width, thickness, and Cross section [mm]. }
\label{tab:dimension}
\centering
\begin{tabular}{cccc}
\toprule
       \textbf{Length} $l_0$ & \textbf{Width} $W$ & \textbf{Thickness} $t$ & \textbf{Cross section} $A_0$  \\
    \midrule     
         $45$ & $5$ & $0.45$ & 2.25 \\
\bottomrule

\end{tabular}
\end{table}

\noindent
The obtained experimental plots shown in Figure \ref{fit:all} depict a quasi-linear stress-strain mechanical response. Consequently, calibrating the uni-axial tests using the incompressible neo-Hookean hyperelastic material model is sufficient for data fitting, for which the constitutive model reads:

\begin{equation}
    \sigma_h= \mu_s(\lambda^2 - \frac{1}{\lambda})
\end{equation}


\noindent 
with $\sigma_h$ is the neo-Hookean engineering stress, and $\mu_s=2C_{10}$ the initial shear modulus. For an in-depth overview of hyperelastic material modeling, the interested reader is referred to \cite{Holzapfel2002}. The detailed steps describing the material calibration using the neo-Hookean hyperelastic material model are given in \cite{doghri2000mechanics, martins2006comparative}.\\

\subsection{Material calibration and curve fitting}
\begin{figure}[H]
\centering
\includegraphics[scale=0.25]{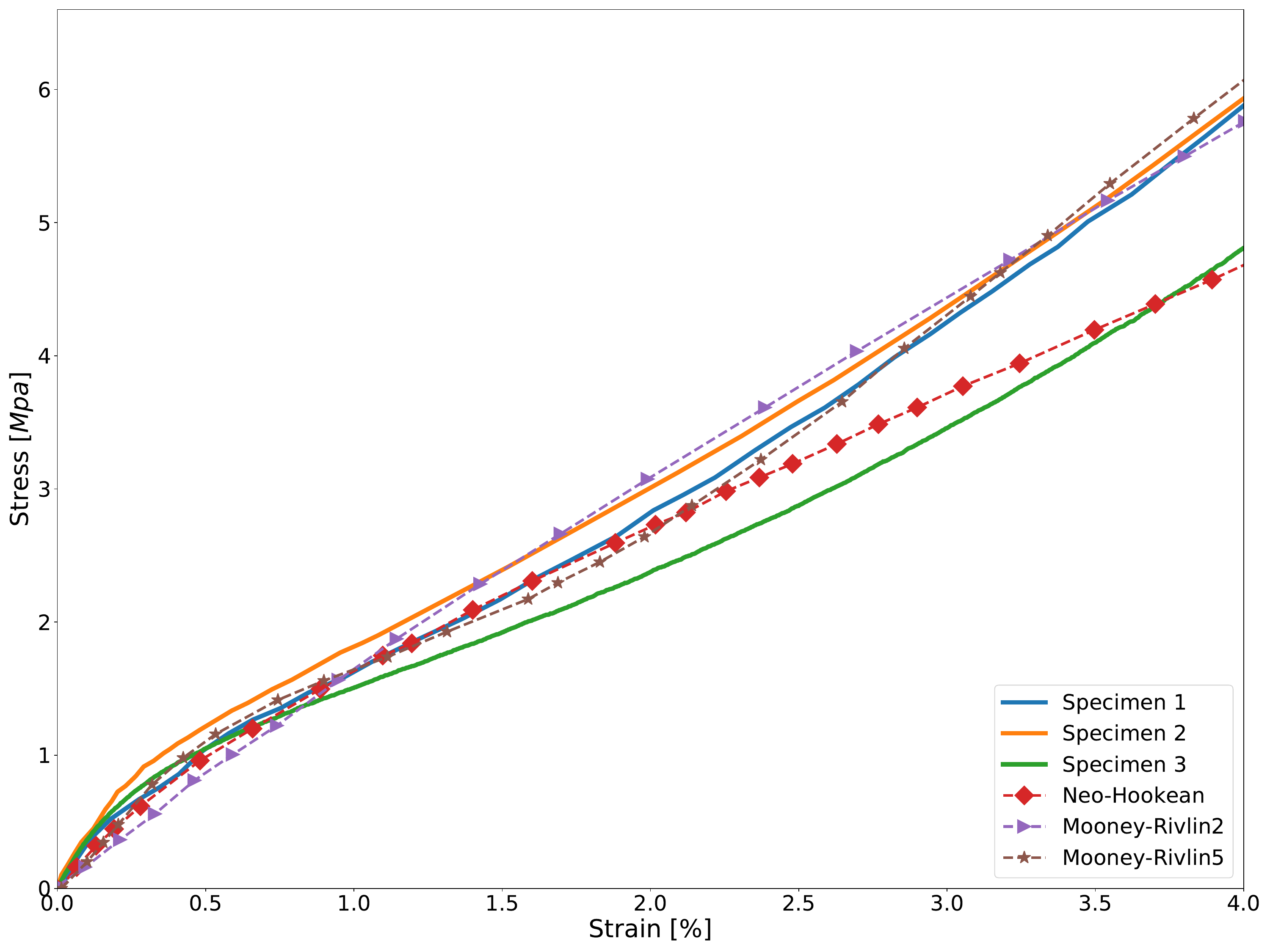}
\caption{Curve-fitting results of the Hyperelastic material calibration. Stress versus strain curve plots for experimental data versus different hyperelastic material models. Red: Curve fitting using the neo-Hookean model. Violet: Curve fitting using the Mooney-Rivlin model with $2$ parameters material model. Cyan: Curve fitting using the Mooney-Rivlin model with $5$ parameters material model. Black: Experimental results.}
\label{fit:all}
\end{figure}

The membrane's material properties play a major role in the numerical analysis of the POCV as the nonlinearity and the expected large deformations emanating from the incoming pressure require a constitutive model to enable accurate replication of the membrane dynamics. As previously mentioned in \ref{constitutive_hyp}, the test results shown in Figure \ref{fit:all} show a quasi-linear trend implying that the neo-Hookean material model would be a convenient choice to calibrate the rubber-like material under study. ANSYS (ANSYS, Canonsburg, PA, USA) supports material calibration using the neo-Hookean material model as well as various other conventional hyperelastic models, such as the Mooney-Rivlin and Ogden models, among others \cite{Martins2006}. A curve-fitting approach is used for material calibration, which incorporates a nonlinear regression analysis for the experimental data as a function of the constitutive hyperelastic model and is described as follows:
\begin{enumerate}
    \item Possession of experimental stress-strain data
    \item Prior knowledge about the structural deformation range expected
    \item Choosing the appropriate hyperelastic material model
    \item Performing a regression analysis using the least squares method
    \item Approximating the material constants associated with the hyperelastic material model from the experimental stress-strain data and constitutive equations
    \item Comparing the experimental data against the obtained results. If the error is large, a more accurate hyperelastic material model could be considered. 
\end{enumerate}
The resulting neo-Hookean stress-strain curve is plotted in Figure \ref{fit:all}, alongside the experimental data. Note that the curve fitting procedure is executed taking into account only sample 1 specimen for simplicity purposes. To further motivate the model selection, the stress versus strain graphs corresponding to two commonly used hyperelastic material models are fitted: The Mooney-Rivlin model with two ($2$) parameters and the Mooney-Rivlin model with five ($5$) parameters.  
The stress-strain plot based on the neo-Hookean model shows good alignment with the experimental data when the specimen's cross-head displacement corresponds to engineering strain values below $100\%$. However, for larger tension strains, higher relative errors are observed. This is expected, as the neo-Hookean constitutive model is better suited for relatively low strain ranges \cite{Holzapfel2002}. Conversely, the results associated with the Money-Rivlin models are more accurate for larger strain ranges. Nevertheless, since the membrane deformations are expected to be below $100\%$ strain ranges, the neo-Hookean model is chosen since it is the simplest hyperelastic constitutive model available and is suitable for capturing the material behavior with sufficient accuracy \cite{Mihai2017}.     

\subsection{Subsystems properties}
The material calibration procedure gives the evaluated parameters in terms of the initial shear modulus $\mu_0$ and the incompressibility parameter $D_0$ with corresponding values equal to $0.946$ MPa, and zero (incompressible material), respectively. The material density is measured experimentally as $1140$ kg/m$^3$. Because of material testing limitations, the material-dependent Rayleigh parameters are borrowed from \cite{Luo2020}. For completeness, the material parameters used in the numerical investigations are summarized in Table~\ref{tab:mat}.

\begin{table}[H]
\caption{Material properties of the fluid and solid used in the simulations. The main and pilot membranes have identical material properties. The contact dummy body material properties are consistent with the material of the POCV casing. The spring is assumed mass-less. The Rayleigh parameters~$\alpha,\beta$ are taken
from~\cite{Luo2020}.}
\label{tab:mat}
\centering
\begin{tabular}{cc}
\toprule
\multicolumn{2}{l}{\textbf{Fluid} (Water)}  \\
\midrule
Density & $1140$ kg/m$^3$ \\

Viscosity & $0.946 \times 10^6$ Pa \\

\toprule
\multicolumn{2}{l}{\textbf{Membranes} (Rubber)}  \\
\midrule
Density & $1140$ Kg/m$^3$ \\

Initial shear stress & $0.946 \times 10^6$ Pa \\

Incompressibility parameter & $0$ \\

$\alpha$ & $0.5$ \\

$\beta$ & $0.005$\\
\toprule
\multicolumn{2}{l}{\textbf{Dummy solid} (Plastic PVC)}  \\
\midrule
Density & $1392$ kg/m$^3$	 \\

Shear Modulus & $1.02 \times 10^9$ Pa \\

Poisson's ratio & $0.4$ \\

Young's Modulus & $2.86\times 10^9$ Pa\\

Bulk Modulus  & $4.78\times10^9$ Pa\\
\toprule
\multicolumn{2}{l}{\textbf{Spring}}  \\
\midrule
 Longitudinal stiffness & $0.1$ N/mm\\
 
 Free length & $14$ mm \\

 Longitudinal damping & $0$ N/mm \\

 Spring length & $4.868 $ mm\\

 Endpoints & Membrane-Fixed reference point\\
\bottomrule

\end{tabular}
\end{table}

\newpage

\section*{Acknowledgments}
The authors wish to express their gratitude to the European Union Horizon 2020 research and innovation program for funding Ahmed Aissa Berraies' Ph.D. fellowship, which is a part of the Marie Skłodowska-Curie ITN-EJD ProTechTion actions (Grant Number 76463). The authors acknowledge also partial financial support from the Ministry of Enterprise and Made in Italy and Lombardy Region through the project “PRotesi innOvaTivE per applicazioni vaScolari ed ortopedIChe e mediante Additive Manufacturing" - CUP : B19J22002460005, on the finance  Asse I, Azione 1.1.3 PON Businesses and Competitiveness 2014 - 2020. The authors are grateful for the computing resources provided by the SurfSara center. Special thanks are given to ANSYS Inc. for their generous contribution in providing the necessary academic and HPC licenses. 


\section*{Data availability}

The model and data generated and analyzed during the present investigation are not publicly available due to commercial sensitivity. However, specific information could be provided by the corresponding author upon a reasonable request.

\section*{Conflict of interest}

The authors declare that they have no known competing interests or personal relationships that could have appeared to influence the work reported in this paper.




 \bibliographystyle{elsarticle-num} 
 \bibliography{manuscript}





\end{document}